**Bone tools, ornaments and other unusual objects during the Middle to Upper Palaeolithic transition in Italy**

Simona Arrighi[a,b*], Adriana Moroni [b], Laura Tassoni[c], Francesco Boschin[b], Federica Badino[a,d], Eugenio Bortolini[a], Paolo Boscato[b], Jacopo Crezzini[b], Carla Figus[a], Manuela Forte[c], Federico Lugli[a,e], Giulia Marciani[a,b], Gregorio Oxilia [a], Fabio Negrino[f], Julien Riel-Salvatore [g], Matteo Romandini[a,c], Enza Elena Spinapolice[h], Marco Peresani[c], Annamaria Ronchitelli[a], Stefano Benazzi[a,i]

*Corrisponding author

a. Università di Bologna, Dipartimento di Beni Culturali. Via degli Ariani 1 48121 Ravenna, Italy. Simona Arrighi: simona.arrighi@unibo.it; Federica Badino: federica.badino@unibo.it; Eugenio Bortolini: eugenio.bortolini2@unibo.it; Carla Figus: carla.figus@unibo.it; Federico Lugli: federico.lugli6@unibo.it; Giulia Marciani: giulia.marciani@unibo.it; Gregorio Oxilia: gregorio.oxilia@unibo.it; Matteo Romandini: matteo.romandini@unibo.it; Stefano Benazzi: stefano.benazzi@unibo.it

b. Dipartimento di Scienze Fisiche, della Terra e dell'Ambiente, U. R. Preistoria e Antropologia, Università di Siena. Via Laterina 8, 53100, Siena, Italy. Adriana Moroni: adriana.moroni@unisi.it; Paolo Boscato: paolo.boscato@unisi.it; Francesco Boschin: fboschin@hotmail.com; Jacopo Crezzini: jacopocrezzini@gmail.com; Annamaria Ronchitelli: annamaria.ronchitelli@unisi.it;

c. Dipartimento di Studi Umanistici, Sezione di Scienze Preistoriche e Antropologiche, Università di Ferrara. Corso Ercole I d'Este 32, 44100, Ferrara, Italy. Laura Tassoni: laura.tassoni@student.unife.it; Manuela Forte: manuelaforte88@gmail.com;Marco Peresani: marco.peresani@unife.it

d. C.N.R. - Istituto per la Dinamica dei Processi Ambientali, Via Mario Bianco 9, 20126 Milano,

e. Dipartimento di Scienze Chimiche e Geologiche, Università di Modena e Reggio Emilia, Via Giuseppe Campi 103, 41125 Modena,

f. Dipartimento di Antichità, Filosofia e Storia, Università degli Studi di Genova. Via Balbi 2, 16126, Genova, Italy. Fabio Negrino: fabio.negrino@unige.it

g. Département d'Anthropologie Université de Montréal, 2900 Boulevard Edouard-Montpetit, Montréal, QC H3T 1J4. Canada. Julien Riel-Salvatore: julien.riel-salvatore@umontreal.ca

h. Dipartimento di Scienze dell'Antichità, Università degli Studi di Roma "La Sapienza". Piazzale Aldo Moro 5, 00185. Roma, Italy. Enza Elena Spinapolice: enzaelena.spinapolice@uniroma1.it

i. Department of Human Evolution, Max Planck Institute for Evolutionary Anthropology, Deutscher Platz 604103 Leipzig

**Abstract**

The arrival of Modern Humans (MHs) in Europe between 50 ka and 36 ka coincides with significant changes in human behaviour, regarding the production of tools, the exploitation of resources and the systematic use of ornaments and colouring substances. The emergence of the so-called modern behaviours is usually associated with MHs, although in these last decades findings relating to symbolic thinking of pre-Sapiens groups have been claimed. In this paper we present a synthesis of the Italian evidence concerning bone manufacturing and the use of ornaments and pigments in the time span encompassing the demise of Neandertals and their replacement by MHs. Current data show that Mousterian bone tools are mostly obtained from bone fragments used "as is". Conversely an organized production of “fine shaped” bone tools is characteristic of the Uluzzian and the Protoaurignacian, when the complexity inherent in the manufacturing processes suggests that bone artefacts are not to be considered as expedient resources. Some traces of symbolic activities are associated to Neandertals in Northern Italy. Ornaments (mostly tusk shells) and pigments used for decorative purposes are well recorded during the Uluzzian. Their features and distribution witness to an intriguing cultural homogeneity within this technocomplex. The Protoaurignacian is characterized by a wider archaeological evidence, consisting of personal ornaments (mostly pierced gastropods), pigments and artistic items.

## 1.Introduction

The dispersal of Modern Humans (MHs) in Europe between 50 ka and 39 ka BP and the concomitant demise of Neandertal populations, are connected to the emergence/diffusion of the so-called "modern behaviour" that is usually considered associated with "advanced" cognitive and technological skills. Concepts like "behavioural modernity", "symbolic thinking", and others like these are particularly delicate topics, especially in the MP-UP transition contexts, where different human species with most probably different (even if not inferior or superior) psychic characteristics and social structures are involved. We agree with other scholars (Eren et al., 2013 and references therein; see also session A21a-Neanderthals on their own terms: new perspectives for the study of Middle Paleolithic behaviour by Chacón M.G. and Rivals F. of the XVII World UISPP Congress Burgos, 1-7 September 2014) in questioning the utility of indistinctly applying these notions, in a comparative way, to extremely varied realities (psychically, geographically, climatically, environmentally, chronologically), as "modernity" and "symbolism" (and their degrees) are in themselves relative and impalpable concepts; therefore inappropriate for scientific comparison. This premise was all the more necessary as our paper focuses on some categories of archaeological materials (i.e. bone tools, ornaments, coloring substances and other things) belonging to a suite of attributes which are generally considered in the literature as proxies of behavioural modernity for their innovative characteristics. In spite of that, as deepening this issue is beyond the scope of our paper, we conformed to current terminology, avoiding, however, too obvious equivalences between symbolic thinking/modern behaviour on the one hand and use of ornaments and non-utilitarian objects in general on the other.

Among technological innovations, bone manufacturing appears to be a pivotal element in outlining what is commonly defined "behavioural modernity", as it entails the occurrence of complex technical skills that make possible the acquisition of an evolved technical system. Complex bone technologies allow the manufacture of functional implements by means of specific competences expressly thought up for bone material, such as scraping, grinding, grooving and polishing (Mellars, 1973; Klein, 1999).These technologies emerge for the first time in Africa during the Middle Stone Age (Brooks et al., 1995; Yellen, 1998; McBrearty and Brooks, 2000; Henshilwood et al., 2001; Jacobs et al.,2006; d'Errico and Henshilwood, 2007; d'Errico et al., 2012a) where they date back to between ~90-60 ka.

In Europe formal bone tools make their appearance in the Middle to Upper Palaeolithic transition complexes, whose attribution to Neandertals or MHs is still controversial (see for example: Benazzi et al., 2011; Mellars, 2011; Hublin et al., 2012; Moroni et al., 2013; Zilhão et al., 2015; Gravina et al., 2018; Moroni et al., 2018), and become a common occurrence in the tool kit of the Upper Palaeolithic groups. During the Middle Palaeolithic Neandertals also produced rare bone scrapers and denticulates (see for example Tromnau, 1983; Hahn, 1976; Freund, 1987), but techniques used in the making of these implements were replicas of lithic knapping, thus attesting the lack of specific technologies conceived for this kind of raw material.

Artistic evidence, ornaments, and the use of pigments play a major role in defining "modern behaviour" because of their supposed symbolic value. From an archaeological perspective, the systematic use of personal ornaments and pigments (possibly also connected to body painting) is the ideal proxy from which a number of behavioural characteristics, involving social relationships within a group (in terms of age, gender, social status etc..) and ethnic identity, can be inferred.

Leaving aside the earliest evidence interpreted as symbolic thinking reported from Indonesia (geometric engravings-500.000 years ago-Jordeens et al., 2014), eastern Africa (coloring substances 400.000- 260.000years BP- Brooks et al., 2018), Morocco, and Israel (possible stone figurines - 500.000-233.000 years ago- Goren-Inbar, 1986 and Kuckenburg, 2001), evidence of symbolic activities (ornaments, pigments, engravings) has been witnessed among MHs in Africa and the Levant between 135 and 70 ka BP (Vanhaeren et al., 2006 and 2013; d'Errico et al., 2008 and 2009; Bar-Yosef et al., 2009), to become usual and widespread in the Upper Palaeolithic.

Some "strange objects", without any apparent functional significance and possibly related to symbolic behavior, have been reported in Europe since the Lower Palaeolithic. They usually are fossils, crystals, stones with cupules, objects resembling anthropomorphic figures, and artifacts made of rare rocks (Moncel et al., 2012). This kind of evidence becomes more frequent in Middle Paleolithic sites and significantly increases in the Upper Palaeolithic. Furthermore, possibly ornamental objects, like claws, feathers, shells, ochre etc, are documented in the European Middle Palaeolithic in general, and appear to be much more common in the Late Mousterian (Zilhão et al., 2010; Peresani et al., 2011; Romandini et al., 2014b).

Symbolic evidence is well documented during the Middle to Upper Palaeolithic transition. Abundant ornaments made on shell, bone, as well as pigments and some engraved items have been retrieved in Chatelperronian and Uluzzian sites. Chatelperronian yielded also teeth pendants. In the Protoaurignacian, such objects are widespread, and personal adornments are particularly common in France, Spain and Italy. They are mostly composed of marine shell beads, usually small gastropods. This paper is intended as a bibliographically informed review of the state-of-the-art account concerning the production of bone tools and ornaments as well as other evidence of possible symbolic thinking (i.e. colouring substances and engraved items) during the time span encompassing the Middle to Upper Palaeolithic transition in Italy (Table 1 and Fig.1), namely the period corresponding to the middle of MIS3 (details on the chronology of MIS3 sites in Italy can be found in Marciani et al., in this special issue). In carrying out this review an attempt was made to match data from unequal bibliographic sources which have offered different degrees of information depending on the ages of publications and the research traditions, the content, level (preliminary exhaustive etc..) and applied methodologies of the studies, with, sometimes, great disparity among the available data. Despite these limits we are convinced that a comprehensive overview of all published materials in the fields of bone industry and of the so-called non-utilitarian objects *sensu lato* can be of great help in addressing future research on the transitional period in Italy. Actually this country plays a key role in the understanding of the dynamics that drove the shift from Neandertals to MHs, due to the presence of a good number of late Mousterian, transitional and early Upper Palaeolithic sites located at very different latitudes, that means different environments and climatic conditions (Badino et al. and Romandini et al. in this special issue). Moreover, Southern Italy documents the earliest evidence of MHs in Mediterranean Europe, offering a preferential point of view to examine this topical chapter of the human evolution. The Italian evidence is discussed considering the coeval European framework, with the aim to highlight those elements which were symptomatic of changes or continuities in human behaviour.

| **Site** | **Techno-complex** | **Osseous tools** | **Ornaments** | **Artistic evidence** | **Pigments** |
|---|---|---|---|---|---|
| **Rio Secco** | Mousterian | • | • | | |
| | | | | | |
| **S. Bernardino** | Mousterian | • | | | |
| | | | | | |
| **R. del Broion** | Uluzzian | • | • | • | • |
| | | | | | |
| **Fumane** | Mousterian | • | • | | |
| | Uluzzian | • | | | |
| | Protoaurignacian | • | • | • | • |
| | | | | | |
| **Tagliente** | Mousterian | • | | | |
| | | | | | |
| **Mochi** | Protoaurignacian | • | • | • | • |
| | | | | | |
| **Bombrini** | Protoaurignacian | • | • | • | • |
| | | | | | |
| **La Fabbrica** | Uluzzian | • | | | |
| | | | | | |
| | | | | | |
| **Reali** | Mousterian | • | | | |
| **Castelcivita** | Mousterian | • | | | |
| | Uluzzian | • | • | | |
| | Protoaurignacian | • | • | | |
| | | | | | |
| **Cala** | Uluzzian | • | • | | |
| | Protoaurignacian | • | • | | |
| | | | | | |
| | | | | | |
| **Paglicci** | Protoaurignacian | • | | | |
| | | | | | |
| **Oscurusciuto** | Mousterian | • | | | |
| | | | | | |
| **Cavallo** | Mousterian | | | ? | |
| | Uluzzian | • | • | | • |
| | | | | | |
| | | | | | |
| **Serra Cicora A** | Protoaurignacian | • | | | • |
| | | | | | |

**Tab. 1:** List of the Italian sites yielding osseous tools, ornaments, artistic evidence and pigments with MIS3 human occupations.

## 2. Bone tools

### *2.1 The Late Mousterian*

Bone tools are mainly represented by the so called "retouchers", usually diaphysis fragments used without any modification to retouch stone tools. Although such implements occurred also previously (i.e., for instance, at Grotta de Nadale - Jéquier et al., 2015- and Grotta Ghiacciaia (Bertola et al., 1999; Thun Hohenstein et al., 2018), they are much more documented during MIS3 at the end of the Mousterian, especially in northern sites (Table 2). Grotta di Fumane (ex Grotta Solinas) and Riparo Tagliente (Monti Lessini) yielded the most abundant evidence. In the very Late Mousterian of Fumane most of the over one hundred of bone retouchers were made on bones of ungulates (mainly Megaceros sp.), and occasionally of bear and ibex (Jéquier et al., 2012 and 2018). Tibiae and femora were the selected anatomical portions, but also metapodials, phalanxes and, exceptionally, canines were used (Fig. 2).

Riparo Tagliente (Monti Lessini) yielded 75 bone retouchers (Leonardi, 1979; Thun Hohenstein et al., 2018). Long bone diaphyses of red deer are the most exploited raw material, but also aurochs/bison and elk bones are present. In the lowermost layers, the exploitation of smaller-sized animals such as roe deer and chamois is recorded as well. Metapodials, tibiae, humeri, radi and femora along with a phalanx and a rib were selected for retouchers. At Rio Secco (Pradis Plateau), 6 retouchers on bone fragments from large mammals (one of which probably an elk, and other four from bear) were retrieved (Peresani et al., 2014a and Romandini et al., 2018). Few bone retouchers (3) have been reported also at Grotta Maggiore di San Bernardino (Colli Berici) (Malerba and Giacobini, 1996).

In Southern Italy there is little evidence of bone retouchers as such implements were found only at Grotta Reali (Molise) (Thun Hohenstein and Bertolini, 2012) and Riparo L'Oscurusciuto (Apulia). The three retouchers found at the former site are obtained from long bone diaphysis (one on a metapodial of red deer). At the Oscurusciuto rockshelter two retouchers (unpublished) were obtained from large size ungulates. As the research stands now, finding an explanation for the scarcity of retouchers in southern sites is quite difficult. The usual presence of concretions on bone findings from most southern sites could be the reason why percussion traces are not easily identifiable.

Exceptionally a "hammer" from a red deer antler was found at Grotta di Castelcivita (Campania) (Gambassini, 1997) and a jaw fragment (probably auroch) with striations comes from Grotta Breuil (Latium) (Alhaique et al., 2006). Furthermore, a retouched bone shaft was retrieved in the Late Mousterian layers of Grotta Fumane (dated to 45-44 ky BP) (Romandini et al., 2014a).

Finally, some so-called "points" from Mousterian layers (Mussi, 1990) were found at Riparo Mochi, Grotta del Broion, Grotta Bernardini (Apulia) and Grotta di Serra Cicora (Apulia); however no taphonomic data are available for these findings, which nature remains, therefore, uncertain.

| | **Fumane** | **Tagliente** | **Rio Secco** | **Grotta Maggiore** | **Reali** | **Oscurusciuto** | **Total** |
|---|---|---|---|---|---|---|---|
| *Ursus arctos* | 1 | | 4 | | | | 5 |
| *Cervus elaphus* | 49 | 7 | | | 1 | | 57 |
| *Alces alces* | | 3 | | | | | 3 |
| *Cervus-Alces-Megaloceros* | 12 | | | | | | 12 |
| *Megaloceros-Alces-Bos/Bison* | 1 | | | | | | 1 |
| *Megaloceros-Bos/Bison* | 3 | | | | | | 3 |
| *Cervidae* | 1 | | | | | | 1 |
| *Bos/Bison* | 2 | 3 | | | | | 5 |
| *Bos-Equus.* | | | | | | 2 | 2 |
| *Rupicapra rupicapra* | 1 | 1 | | | | | 2 |
| *Capra ibex* | 1 | | | | | | 1 |
| *Capreolus capreolus* | | 1 | | | | | 1 |
| *Ungulata* | 2 | 48 | | | | | 50 |
| Unid. big size | 7 | | 2 | | 2 | | 11 |
| Unid. | | | | 3 | | | 3 |
| **Total** | **80** | **63** | **6** | **3** | **3** | **2** | **157** |

**Table 2**: Taxonomical attribution of the retouchers recovered in Late Mousterian (<50ka) sites in Italy (data from: Malerba and Gicobini, 1996; Jéquier et al., 2012; Thun Hohenstein and Bertolini, 2012; Peresani et al., 2014; Romandini et al., 2018; Thun Hohenstein et al., 2018)

*2.1 The Uluzzian*

The Uluzzian bone kit displays substantial changes with respect to the Mousterian, especially due to the emergence of formal tools. The technological process by which bone tools were produced appears to be part of a tradition shared by all the Uluzzian groups in Italy, with the only exception of Fumane (Table 3). This process entails obtaining awls and cylinder – conical elements, from specific anatomical parts, like metapodials of red deer and fibulae and metapodials of horse.
The sites which yielded the greater part of bone tools are in Southern Italy. At Grotta del Cavallo (Apulia) 8 specimens (Fig. 3, 1-8), mostly awls or fragmentary awls, were retrieved (Palma di Cesnola, 1966; d’Errico et al., 2012b). Grotta di Castelcivita (Campania) returned 6 pieces: four awls (Fig. 3, 10-12, 14), one point and a double pointed implement (Fig.3, 14), interpreted as a hook (d’Errico et al., 2012b). A single awl (Fig. 3, 9) has been found at Grotta della Cala (d’Errico et al., 2012b). A sole specimen (Fig. 3, 15) is known in Central Italy from Grotta La Fabbrica (Tuscany) (Pitti et al., 1976; Villa et al., 2018); it shows a coating of ochre at its base and other residual traces along its shaft. Villa et al. (2018) draw a parallel between this object and the ochred bone tools of the Still Bay phase from Blombos Cave, dated to 75–72 ka (Henshilwood et al., 2001; Henshilwood, 2012).
According to d’Errico and colleagues (2012b), the Uluzzian bone tools were produced using at least three different techniques: scraping the end of naturally pointed elements, modifying thin lengthened shaft fragments, shaping elongated epidiaphyseal fragments. The same authors argue that these implements were utilized to perforate a range of materials from relatively hard (Cavallo and Cala), like thick leather, to soft like skin, furs and vegetal substances (Castelcivita). Noteworthy is the retrieval of a splintered piece on bone at Grotta del Cavallo (Borgia et al., 2017).
In Northern Italy Riparo Broion yielded 4 artefacts: 3 awls and pointed tools (Fig. 3, 16-18) and a probable needle (Fig. 3, 19) The exploitation of anatomical elements naturally shaped *ad hoc* (ulna or telemetacarpals) and their subsequent shaping by mean of a lithic tool is attested at least for two of these implements. The other tools, obtained from unidentified skeletal portions, were shaped by longitudinal scraping. Polished areas were identified on 3 pointed tools, but a dedicate functional study was not carried out (Peresani et al., 2019a).
The Uluzzian bone kit from Fumane (layer A3) includes an awl (Fig. 3, 20) and a worked implement (Fig. 3, 21). Unlike the rest of the Uluzzian findings, the skeletal portions exploited at Fumane are exclusively ribs (Peresani et al., 2016). The awl was obtained by longitudinally incising and splitting a medium-large mammal bone and later shaping the distal end by scraping. Use-wear analysis suggests this tool has been used for perforating ochred leather. The second worked object is made from mid-lateral posterior rib portion of a medium-large mammal, with one side shaped by scraping with the aim to obtain a beveled edge.
At Fumane the use of bone retouchers is documented also during the Uluzzian (Jéquier et al., 2012).

| | | **Cavallo** | **Castelcivita** | **Cala** | **Fabbrica** | **Fumane** | **Broion** |
|---|---|---|---|---|---|---|---|
| Species | *E. ferus* | 3 | 1 | | 1 | | |
| | *C. elaphus* | 2 | | 1 | | | |
| | Medium-large size mammals | | | | | 2 | |
| | Unidentified | 3 | 5 | | | | 4 |
| Anatomical portion | Metapodial | 2 | 1 | | 1 | | |
| | Metatarsal | 3 | | 1 | | | |
| | Ulna/Telemetapodial | | | | | | 2 |
| | Fibula | 1 | | | | | |
| | Rib | | | | | 2 | |
| | Unidentified | 2 | 5 | | | | 2 |
| Tool type | Awls | 7 | 2 | | 1 | 1 | 3 |
| | Needles | | | | | | 1 |
| | Unidentified | 1 | 4 | 1 | | 1 | |
| Use-wear | | 3 | 3 | ? | ? ochre | 1 | 3? |

**Table 3**: Archaeozoological and typological classification of the Uluzzian bone tools recovered in Italy (data from: d'Errico et al., 2012b; Peresani et al., 2016; Villa et al., 2018; Peresani et al., 2019a)

*2.3 The Protoaurignacian*

Differences can be stressed between the Protoaurignacian and the Uluzzian in the bone kit. In the North, bone manufacturing acquires more technical complexity and there is an increase in the number and typology of bone tools (Table 4), at least in some sites like Fumane. This cave (Bertola et al., 2013) yielded from the oldest Protoaurignacian complex (layers A2 and A1) a rich assemblage, composed of points and awls, often showing fractures due to their use (Fig. 4. 9-10). These pieces were mainly obtained from cervid diaphyses (but in some cases also ribs). Manufacture involved only the shaping of the functional part, according to a process called *poinçon d'économie*. Use-wear analysis suggests their use for piercing. Some specimens, displaying a very thin (less than 5 mm) and elongated point, were probably used as needles. A split-based point (Fig. 4.8) has been retrieved at the interface between layers A1 (Protoaurignacian) and D3 (Aurignacian *sensu lato*) (Bertola et al., 2013). Furthermore, some bone retouchers and smoothers were found (Jéquier, 2014; Jéquier et al., 2018).

Bone tools from Riparo Bombrini (layers A3-A1) (Liguria) (Bertola et al., 2013; Holt et al., 2018) include eight artefacts, mostly fragmentary (Fig. 4. 1-7). They are points (1), needles (2), awls (3) and unidentified fragments (2). A bevelled tool obtained from cervid antler previously ascribed to the Protoaurignacian and coming from a disturbed area, has been dated to the Epigravettian (Holt et al., 2018). The points were made on blanks obtained by indirect percussion, then shaped by scraping. The presence of specific traces suggests that these tools were hafted. Diaphyseal fragments obtained by direct percussion from long bones of large size ungulate had been used as blanks for the awls, which were later shaped only on the distal part, like the ones from Fumane. The needles were obtained from elongated blanks, carefully scraped on the whole surfaces. Use –wear analysis indicates their use on hide.

Riparo Mochi (layer G) (Liguria) yielded pointed artefacts (7), probable awls, points and needles (Kuhn and Stiner, 1992, 1998). The presence of waste products testifies to a local manufacturing of these tools. Also, two antler split-based points had been ascribed to the Protoaurignacian (Kuhn and Stiner, 1992, 1998), but a recent reassessment of the stratigraphic provenance of these findings led to their re-attribution to the Aurignacian rather than to the Protoaurignacian (Tejero and Grimaldi, 2015).

Contrary to the north, bone implements are rare and scarcely diversified in Southern Italy, where it is the Uluzzian rather than the Protoaurignacian to have returned the majority of these artifacts. Few Protoaurignacian specimens (Table 4) are documented at Grotta di Castelcivita (an awl from a metapodial of roe deer) (Gambassini, 1997), Grotta della Cala (four fragmentary bone points of which one is from a rib) (Benini et al., 1997; Fig. 4.13-16) and Grotta Paglicci (an awl made on a shaft fragment of a medium-large mammal) (Borgia et al., 2016; Fig. 4.12). A fragmented awl has been also recorded in the so-called "Uluzzo-Aurignaziano" of Serra Cicora A (Spennato, 1981; Palma di Cesnola, 1993).

| | **Fumane** | **Mochi** | **Bombrini** | **Castelcivita** | **Cala** | **Paglicci** | **Serra Cicora A** |
|---|---|---|---|---|---|---|---|
| Awls | 23 (3) | | 4 | 1 | | 1 | 1 |
| Needles | | | 2 | | | | |
| Points | (19) | | | | 3 | | |
| Unidentified pointed tools | | 7 | 1 | | | | |
| Smoothers | 2 | | | | | | |
| Unidentified | | | 1 | | | | |
| **Total** | **54 (22)** | **7** | **8** | **1** | **3** | **1** | **1** |

**Table 3:** Protoaurignacian bone tool kits (data from: Spennato, 1981; Benini et al., 1997; Gambassini, 1997; Kuhn and Stiner, 1998; Bertola et al., 2013; Jéquier, 2014; Tejero and Grimaldi, 2015; Borgia et al., 2016)

## 3 Ornaments and other unusual objects

### *3.1 The late Mousterian*

The earliest evidence connected to symbolic behaviour can be found in the Mousterian. The exploitation of animal resources for ornamental purposes has been reported in the North. Eagle claws and raptor bones with cut-marks (Fig. 5), indicating the intentional removals of the claws themselves and of the flight (remex) feathers, were found at the caves of Rio Secco (Romandini et al., 2014b) and Fumane (Peresani et al., 2011). Both the claws and the feathers have been interpreted by the authors as ornamental items. Layer A9 of Fumane, dated to 47.6 cal ka BP (minimum age), yielded the only italian Mousterian shell (Fig.5) interpreted as an exotic object, colored with red ocher and suspended by a “thread” for visual display as a pendant (Peresani et al., 2013). This is a fossil marine shell, *Aspa marginata*, probably collected in Miocene or Pliocene fossil outcrops located, at least, about one hundred kilometers south of the site.
In addition to the above-mentioned evidence, the Mousterian in Italy is characterized by the occurrence of engraved bones and stones bearing linear signs. This is the case of several objects retrieved at Riparo Tagliente, Grotta S. Bernardino, Grotta di Fumane (ex Grotta Solinas) in Veneto which were described in old publications (Leonardi, 1975, 1980, 1983, 1988). Despite their anthropogenic origin, there is no evidence attesting to a symbolic meaning of these engravings (Peresani et al, 2014b). At Grotta Costantini (Liguria) there is a horse rib showing three groups of linear marks (Bachechi, 2001), reported as coming from the top of the Mousterian but more likely it is an intrusive Upper Palaeolithic artefact. In Central Italy, objects with linear incisions are present in the Middle Palaeolithic contexts of Grotta di Gosto (Tuscany) (Tozzi, 1974 but see Moroni et al., 2018 for doubts on the chronology of this site) and Valle Radice (Latium) (Biddittu et al., 1967). In Southern Italy, Grotta del Cavallo (Martini et al., 2004) and Grotta dell'Alto (Borzatti von Löwestern and Magaldi, 1967) yielded four and two engraved stones respectively.

### *3.2 The Uluzzian*

In the Uluzzian the use of adornment objects is systematic, thus revealing the emergence of a well-established tradition related to non-utilitarian activities. The use of shell beads, usually from tusk specimens (*Antalis* sp.), for ornamental purposes is broadly attested in Uluzzian sites (Table 5). Grotta del Cavallo yielded the largest ornamental assemblage (a few hundred), composed mainly of tusk shells, from the whole Uluzzian sequence. Gastropods are fewer in number and overall occur in the final phase (Palma di Cesnola, 1993). Also some fragmentary bivalves are recorded. A number of marine shell beads along with a coral branch were found in the Uluzzian layer of Grotta della Cala (Ronchitelli et al., 2009). The perforated (twenty-four scaphopods, six gastropods and two *Glycimeris nummaria*- syn. *G. insubrica*) and non-perforated marine molluscs of this site includes 78 items (Fiocchi, 1998; Ronchitelli et al., 2009). Several marine shells (gastropods and bivalves) are also documented in the Uluzzian layers of Castelcivita, but none of them show any kind of perforation (Gambassini, 1997).

Riparo Broion yielded five worn tusk beads and a pierced freshwater gastropod (*Theodoxus danubialis*) (Peresani et al., 2019a).

Colouring materials (lumps of ochre and limonite) were recovered at Grotta del Cavallo and Grotta Mario Bernardini (Palma di Cesnola, 1989)**.** Some oxidized glomeruli were found at Grotta della Cala. Traces of ochre have been identified on two tusk beads from Riparo Broion (Peresani et al., in 2019a) and on several stone tools from the whole Uluzzian package of Grotta del Cavallo (Moroni et al., 2018)

| **Class** | **Family** | **Species** | **Cavallo*** | **Cala*** | **Broion** | **Total** |
|---|---|---|---|---|---|---|
| Gastropoda | Unidentified | | | • | | |
| | Archeo gastropoda unidentified | | | • | | |
| | Haliotidae | *Haliotis tuberculata lamellosa* | | • | | |
| | Trochidae | *Clanculus* sp. | | • | | |
| | Coloniidae | *Homalopoma sanguineum* | | • | | |
| | Cerithiidae | Unidentified genus | | • | | |
| | Triviidae | *Trivia mediterranea* (syn: *T. pulex*) | | • | | |
| | | *Trivia* sp. | | • | | |
| | Naticidae | cf. Naticidae gen. sp. unidentified | | • | | |
| | Cassidae | *Galeodea echinophora* | | • | | |
| | | cf. *Galeodea* sp. | | • | | |
| | Ceritithiopsidae | Unidentified genus | | • | | |

| | | | | | |
|---|---|---|---|---|---|
| | Nassaridae | *Tritia incrassata* (syn: *Nassarius incrassatus*) | | • | |
| | | *Tritia neritea* (syn: *Cyclope neritea)* | • | | |
| | | *Tritia pellucida* (syn: *Cyclope pellucida*) | | • | |
| | Columbellidae | *Columbella rustica* | • | | |
| | Neritidae (Freshwater) | *Theodoxus danubialis* | | | • |
| Bivalvia | Unidentified | | | • | |
| | Glycymerididae | *Glycymeris nummaria* (syn: *G. insubrica*) | | • | |
| | Pectinidae | Pectinidae sp. Gen. unidentified | | • | |
| | | *Pecten jacobeus* | | • | |
| | Veneridae | Veneridae unidentified (cfr. *Callista chione*) | | • | |
| | | *Callista chione* | | • | |
| Scaphopoda | Dentaliidae | *Antalis dentalis/inaequicostata* | • | • | • |
| | | *Antalis vulgaris* | • | • | • |
| | | *Antalis* cfr. *vulgaris* | | • | |
| | | *Antalis* sp. | | • | |
| | Fustiariidae | *Fustiaria rubescens* | | • | |
| Unidentified | | | | • | |

**Table 5:** Ornamental shell taxa (pierced and not pierced) from Uluzzian sites (data from Palma di Cesnola, 1993; Fiocchi, 1998; Ronchitelli et al., 2009, Peresani et al., 2019a). Classification and nomenclature used for molluscs is based on the systematics index of S.I.M. - Società Italiana di Malacologia (www.societaitalianadimalacologia.it) and WoRMS - World Register of Marine Species (www.marinespecies.org). * The revision of the ornamental shell assemblages from Grotta del Cavallo and Grotta della Cala is currently ongoing, therefore data presented here are borrowed from previous publications.

### *3.3 The Protoaurignacian*

The Protoaurignacian is characterized by a wider range of personal adornments, mainly consisting of marine shell beads (Tables 6 and 7). In comparison with the Uluzzian, the number of retrieved elements considerably increases as does the number of the sites where they were found (Riparo Mochi, Riparo Bombrini, Grotta di Fumane - layers A2 and A1 -, Grotta della Cala and Grotta di Castelcivita) (Fig.7). Ornamental species mostly include gastropods (e.g. *Tritia* sp. and other Nassaridae, *Homalopoma sanguineum* and Trochidae) and to a lesser extent, bivalves (e.g. *Glycymeris nummaria* and other Glycymerididae, *Acanthocardia tuberculata* and other Cardiidae). Tusk specimens are generally very few in number. Shell assemblages from coastal sites (Mochi,

Bombrini, Cala) do not show any selective choice regarding a particular species (Fiocchi, 1998), even if a remarkable presence of *H. sanguineum* could be noted. At Fumane, which was located at about 200 km from the Tyrrhenian coast and 400 km from the Adriatic one, there was, on the contrary, a selection in favour of externally red coloured species: *Homalopoma sanguineum*, *Clanculus corallinus* and *Clanculus cruciatus* (Peresani et al., 2019b). The taxa spectrum of the ornamental shells of Fumane shows analogies with the Ligurian sites, supporting evidence for contacts between the two areas. Anyway, an Adriatic provenance of these shells has also been envisaged (Bertola et al., 2013).

Interestingly, only seashells seem to have been used as ornaments in Southern Italy, whereas, in the North, these are often associated with bone and stone ornaments. Teeth pendants have been recovered at Grotta di Fumane and stone ornaments occur at Bombrini (Fig 7). This site yielded belemnite pendants and soapstone artefacts, among which a perforated one. The provenance of the soapstone collected at Bombrini is almost certainly from the Apennine chain between Liguria and Emilia (Chella, 2002; Gernone and Maggi, 1998; Negrino et al., 2017: Holt et al., 2018). The non-shell ornaments from Mochi - three beads made on soapstone, resembling *craches* of red deer, two pendants made on fossil belemnite, an ivory basket-shaped bead and two teeth pendants - previously ascribed to the Protoaurignacian, have been recently attributed to the Early Aurignacian (Tejero and Grimaldi, 2015).

"Artistic" evidence is extremely poor in the Protoaurignacian and limited to Northern Italy. This consists of notch and incision patterns on bones from Riparo Bombrini, Riparo Mochi and Grotta di Fumane (Fig.7). Ochre is well documented in the Northern sites (Kuhn and Stiner, 1998; Bietti and Negrino, 2008; Cavallo et al., 2017) with evidence of heating treatment at Fumane (Cavallo et al., 2018).

| | **Fumane** | **Bombrini** | **Mochi** | **Castelcivita** | **Cala** | **Serra Cicora A** |
|---|---|---|---|---|---|---|
| Shell ornaments | • | • | • | • | • | |
| Bone ornaments | • | • | | | | |
| Stone ornaments | | • | | | | |
| Pigments | • | • | • | | | • |
| Artistic evidence | • | • | • | | | |

**Table 6:** Presence of non-utilitarian items in Protoaurignacian sites of Italy

| **Class** | **Family** | **Species** | **Fumane** | **Bombrini** | **Mochi** | **Cala** | **Castelcivita** |
|---|---|---|---|---|---|---|---|
| Gastropoda | Patellidae | *Patella* cfr. *ulyssiponensis* | • | | | | |
| | Fissurellidae | *Fissurella* sp. | | | • | | |
| | Haliotidae | *Haliotis tuberculata lamellosa* | | | • | • | |
| | Trochidae | *Gibbula albida* | | | | • | |
| | | *Gibbula ardens* | | | | • | |
| | | *Gibbula turbinoides* | • | | | | |
| | | *Gibbula* sp. | • | | • | • | |
| | | *Steromphala adansonii* (syn: *Gibbula adonsonii*) | • | • | | • | |
| | | *Steromphala varia* | • | | | | |
| | | *Jujubinus striatus* | | | | • | |
| | | *Jujubinus* sp. | • | | • | • | |
| | | *Phorcus articulatus* (syn: *Osilinus articulates*) | • | • | | • | |
| | | *Phorcus richardi* (syn: *Gibbula richardi*) | • | | | | |
| | | *Phorcus turbinatus* (syn: *Osilinus turbinatus*) | | | | • | |
| | | *Phorcus* sp. | | | | • | |
| | | *Clanculus corallinus* | • | | | • | |
| | | *Clanculus cruciatus* | • | | | • | |
| | | *Clanculus jussieui* | • | • | | • | |
| | | *Clanculus* sp. | • | • | • | • | |
| | | Unidentified genus | • | • | • | • | |
| | Turbinidae | *Bolma rugosa* | | | • | • | |
| | Coloniidae | *Homalopoma sanguineum* | • | • | • | • | • |
| | Phasianellidae | *Tricolia pullus* | • | | | • | |
| | | *Tricolia speciosa* | | | | • | |
| | Cerithiidae | *Bittium latreillii* | • | | | | |
| | | *Bittium reticulatum* | • | | | • | |
| | | *Cerithium vulgatum* | • | | | • | |
| | | *Cerithium* sp. | | | • | • | |
| | Cerithiopsidae | *Cerithiopsis* sp. | • | | | | |
| | Turritellidae | *Turritella communis* | • | • | | | |
| | | *Turritella* sp. | | | • | | |
| | Littorinidae | *Littorina obtusata* | • | | | | |
| | | *Littorina saxatilis* | • | | | | |
| | | *Littorina* sp. | | | • | | |
| | | *Melaraphe neritoides* (syn: *Littorina neritoides*) | | | | | • |
| | Rissoidae | *Rissoa variabilis* | • | | | | |
| | | *Rissoa* sp. | • | | | | |
| | Aporrhaiidae | *Aporrhais pespelecani* | | • | • | | |
| | Triviidae | *Trivia arctica* | • | • | | • | |
| | | *Trivia mediterranea* (syn: *T. pulex*) | | | | • | |
| | | *Trivia* sp. | | | • | • | |
| | Cypraeidae | *Luria lurida* | • | | | | |
| | | *Luria* sp. | • | | • | | |
| | Naticidae | *Euspira macilenta* | • | | | | |
| | | *Naticarius hebraeus* (syn: *Natica hebraea*) | | | | • | |
| | | *Euspira* sp. | • | | | | |
| | | Unidentified genus | • | | • | • | |
| | Cassidae | *Semicassis saburon* (syn: *Phalium saburon*) | | | | • | |
| | Muricidae | *Ocenebra edwardsii* | • | • | • | | |
| | | *Ocinebrina* sp. | • | | | | |
| | | Unidentified genus | • | | | | |
| | Mitridae | *Episcomitra cornicula* (syn: *Mitra cornicula*) | • | | | • | |
| | | Unidentified genus | | | | • | |
| | Buccinidae | *Aplus* sp. | • | | | | |
| | | Unidentified genus | • | | | | |
| | Pisaniidae | *Gemophos viverratoides* (syn: *Pollia viverratoides*) | | | | • | |
| | Nassaridae | *Nassarius circumcinctus* | • | • | | • | |
| | | *Tritia corniculum* (syn: *Nassarius corniculum*) | • | | | | |
| | | *Nassarius gibbosulus* | | • | | | |
| | | *Tritia cuvierii* (syn: *Nassarius costulatus cuvierii*) | | | | • | |
| | | *Tritia incrassata* (syn: *Nassarius incrassatus*) | • | | | • | |
| | | *Tritia mutabilis* (syn: *Nassarius mutabilis*) | • | | • | • | |
| | | *Tritia neritea (*syn: *Cyclope neritea)* | • | | | | |
| | | *Tritia pellucida (*syn: *Tritia pellucida)* | • | | | | |
| | | *Tritia reticulata* (syn: *Nassarius reticulatus*) | • | • | | | |
| | | *Tritia* sp. (syn: *Cyclope* sp.) | • | • | • | • | • |
| | | *Nassarius* sp. | | | • | • | |
| | Columbellidae | *Columbella rustica* | | | | • | |
| | | *Mitrella gervilii* | | | | • | |
| | | *Mitrella scripta* | | | • | | |
| | Cancellariidae | *Bivetiella cancellata* (syn: *Cancellaria cancellata*) | | • | | | |
| | Conidae | *Conus ventricosus* (syn: *C. mediterraneus*) | | | • | • | |
| | Neritidae (Freshwater) | *Theodoxus* cfr. *danubialis* | • | | | | |
| | Unidentified | | • | | | • | |
| Bivalvia | Noetidae | *Striarca lactea* | | | | • | |
| | Glycymerididae | *Glycimeris glycimeris* | • | | | • | • |
| | | *Glycymeris nummaria* (syn: *G. insubrica/ G. violacescens*) | • | | | | |
| | | *Glycymeris* sp. | • | | | • | |
| | Mytilidae | *Mytilus galloprovencialis* | • | | | | |
| | | *Mytilus* sp. | | | | • | |
| | Ostreidae | Unidentified genus | • | | | | |
| | Pectinidae | *Pecten jacobeus* | | | | • | • |
| | | *Chlamys* sp. | | | • | | |

| | | | | | | |
|---|---|---|---|---|---|---|
| | | Unidentified genus | | | • | |
| | Cardiidae | *Acanthocardia tuberculata* | | | • | |
| | | *Cerastoderma glaucum* | • | | • | |
| | | *Cerastoerma* cfr. *edule* | • | | | |
| | | *Cerastoderma* sp. | • | | | |
| | | *Papillicardium papillosum* | • | | | |
| | | Unidentified genus | | | • | |
| | Veneridae | *Callista chione* | | | • | |
| | Unidentified | | | | • | • |
| Scaphopoda | Dentaliidae | *Antalis inaequicostatum* | • | • | • | |
| | Fustiariidae | *Fustiaria rubescens* | | | • | |

**Table 7:** Ornamental shell taxa (pierced and not pierced) found in the Protoaurignacian layers of Italian sites (Modified after Bertola et al., 2013; data from Barge, 1983; Gambassini, 1997; Fiocchi, 1998; Stiner, 1999; Holt et al., 2018, Peresani et al., 2019b). Classification and nomenclature used for molluscs is based on the systematics index of S.I.M. – Società Italiana di Malacologia (www.societaitalianadimalacologia.it) and WoRMS - World Register of Marine Species (www.marinespecies.org).

## 4. Discussion

4.1 *Bone artefacts*

In the Mousterian world in Italy bone was, indeed, an optional raw material, opportunistically exploited for tools that do not result from a planned sequence of actions. These tools mainly consist of unmodified long bone fragments of medium or large ungulates used as retouchers, a kind of implement which has been ascertained to have been in existence in Europe since the Lower Palaeolithic (Blasco et al., 2013; Serangeli et al., 2015; van Kolfschoten et al., 2015; Moigne et al., 2016). The occasional exploitation of other taxa (including humans) is also recorded (Daujear et al., 2014; Jéquier et al., 2012 and 2016; Rougier et al., 2016). Specific studies on the function of Middle Palaeolithic retouchers have demonstrated their main implication in sharpening, blunting, shaping, and crushing cutting-edges of stone tools (Siret, 1925; Vincent, 1993; Armand and Delagnes, 1998; Daujear et al., 2014), thus confirming the “retoucher” nomenclature. Nevertheless, their sporadic use also in different stages of the lithic production (therefore described as hammers or anvils) is admitted (Armand and Delagnes, 1998; Rigaud, 2007; Daujear et al., 2014). The long life of bone retouchers attests to a continuity of this technological tradition encompassing several tens of millennia which extends until the Early Upper Palaeolithic based on the Fumane sequence, where also the Protoaurignacian assemblages have produced similar items (Jéquier et al., 2018). The occurrence of formal bone tools in the Mousterian is a matter of debate since decades (Villa and d’Errico, 2001). However recent excavations in two Mousterian of Acheulian Tradition sites, Pech-de-l’Azé I and Abri Peyrony (Soressi et al., 2013), have brought to light four smoothers (or *lissoirs*) intentionally shaped by polishing. This has opened further speculations on this topic. The discovery of these objects, although representing an exception in the repetitive world of Mousterian retouchers, indicates that the use of a technology like polishing in bone processing was not foreign to Neandertals' nature. Furthermore, a basic shaping or preparation has been also supposed for some bone tools from the Lower Palaeolithic site of Schöningen (Julien et al., 2015)..

The appearance of the transitional techno-complexes is characterised by the systematic production of formal tools. In Europe bone manufacturing is well documented especially in the Châtelperronian, where the greatest amount of formal tools (awls, points and fragmentary items) is provided by the archaeological deposit of Grotte du Renne (France) (d'Errico et al., 1998), whose integrity has been, however, questioned by several authors (White, 2001; Higham et al., 2010; Bar-Yosef and Bordes, 2010; Higham et al., 2011, but see contra: Caron et al., 2011; Hublin et al., 2012). Bone tools also occur in other European Initial Upper Palaeolithic and transitional techno-complexes such as the Bachokirian and the Szeletzian (Churchill and Smith, 2000; Glen and Kaczanowsky, 1982).

In the Uluzzian, awls and cylinder-conical elements are consistently typical formal tools, namely implements designed for specific tasks (usually for piercing soft materials) within the site economy. Even if the technical scheme in processing these tools was quite simple, the whole procedure devoted to their making implied an investment in time and energy for the selection and processing (i.e. disarticulating and defleshing) of suitable anatomical parts (whatever they were) from specific taxa. This selective attention for well definite anatomical parts to be used as blanks is a fundamental step in the *chaine opératoire*, more challenging, in some way, than the production of bone blanks by other methods (e.g. percussion).

In the Protoaurignacian a systematic and sometimes abundant production of bone tools has been reported from several sites all over Europe (mainly from France and Spain), like Trou de la mère Clochette, Grotte du Renne and Isturitz (Julien et al., 2002; Soulier et al., 2014; Tartar, 2015). The sporadic presence in Protoaurignacian assemblages (Trou de la mère Clochette, Arbreda, Fumane) of the split-based points (Broglio et al., 1996; Ortega Cobos et al., 2015; Tartar, 2015) characteristic of the Aurignacian, has been interpreted by some authors (Teyssandier and Liolios, 2008; Tartar, 2015) as being symptomatic of a gradual process leading from the Protoaurignacian to the Aurignacian.

In Italy the Protoaurignacian bone technology does not denote substantial changes if compared to the one of the Uluzzian, except, perhaps, for the selected skeletal portions. In most cases, bone tools imply simple manufacturing mainly aimed to obtain an active part from bone fragments (Fumane). At Riparo Bombrini there is evidence of the production of blanks through direct percussion. Although Protoaurignacian implements are typologically more various (also including points and needles), we cannot identify a genuine hiatus between this techno-complex and the Uluzzian. The occurrence of a real rupture, marked by significant innovations in bone technology and typology, has been postulated by some authors for the Early Aurignacian, when the use of antler was introduced, and bone was also used to produce hunting weapons like the split-based points (Tejero, 2014; Tejero and Grimaldi, 2015). Conversely, bone manufacturing has been interpreted as a subsistence activity closely related to specific domestic occupations both in the Uluzzian and the Protoaurignacian (d'Errico et al., 2012b; Peresani et al, 2016; Bertola et al., 2013). Despite this functional homogeneity, we note that the northern Protoaurignacian displays a preference for blanks from generic diaphyseal parts, contrary to the Uluzzian which is usually more selective and oriented to exploit specific anatomical portions. In the south the poor evidence recorded in Protoaurignacian sites (Grotta di Castelcivita, Grotta della Cala, Paglicci and Serra Cicora) does not allow us to enucleate possible differences with the Uluzzian.

4.2 *Ornaments and other non-utilitarian evidence*

Probable evidence of non-utilitarian activities, such as the occurrence of unusual objects most likely intentionally collected by hominids, has been recognized in Europe since the Lower Palaeolithic. It is difficult to evaluate the meaning of such collecting, but hominids' curiosity for unfamiliar and bizarre objects may have played an important role (Leroi-Gourhan, 1961 and 1964). In Europe the earliest records interpreted as "symbolic" evidence are the grooved bones from Bilzingsleben (Germany) dated to 350-220.000 years ago (Mania and Mania, 1988). Engraved stones and bones have been found in Europe both in Lower and Middle Palaeolithic sites. As a non-exhaustive example, we can recall the objects from Saint Anne I (Raynal and Séguy, 1986; Crémades, 1996) in France, Oldisleben (Bednarik, 2006), and Whylen (Moog, 1939) in Germany. A list of Lower - Middle Palaeolithic engraved stones is reported by Majkić et al. (2018a). Some Authors (Marshack, 1976; Bednarik, 1995; Bahn, 1996) have interpreted these objects as non-utilitarian expressions by Neandertal or Pre-Neandertal hominids, whereas other scholars suggest a more prosaic function, at least for some of them, assuming that they might be related to butchering practices, or even to carnivore activities or other taphonomic phenomena (Bordes, 1969; Raynal, Séguy, 1986; Crémades, 1996; Wolpoff, 1996; d'Errico and Villa, 1997; Majkić et al., 2017). Pigments are also frequently recorded in Mousterian sites (as for example Pech-de l'Azé I); however, their exploitation for non-subsistence activities has been questioned in South African Middle Stone age (Wadley, 2003 and 2005; Dayet et al., 2015).
During the Late Mousterian, discoveries of objects interpreted as non-utilitarian are more frequent. Recent research has highlighted the presence among Neandertals of ochre, unusual lithic objects, mobiliary and cave art as well as claws, feathers and shells, possibly used with ornamental purposes. The intentional removal of raptor claws is documented in France (Morin and Laroulandie, 2012) and Italy (Romandini et al., 2014b). The same evidence has been found in the site of Krapina in Croatia, dated to 130 ka BP (Radovčić et al., 2015). The probable use of naturally pierced bivalves as ornaments and the use of pigments are documented in two Mousterian sites in Spain, Cueva Antón and Cueva de los Aviones (Zilhão et al., 2010) recently re-dated to120-115 ka (Hoffmann et al., 2018a). Furthermore, some cave paintings in Spain (La Pasiega, Maltravieso, Ardales) have been re-dated to 60 (Hoffmann et al., 2018b) or to 47 ka (Slimak et al., 2018) allowing their possibly assignment to Neandertals like the deeply engraved lines in a hash-marked pattern on the bedrock of Gorham's Cave at Gibraltar (Rodríguez-Vidal et al., 2014). Bones showing notches and incisions are also documented in Europe, as, for example, a schematically engraved bone found in the Final Mousterian layer of Bacho Kiro (Bulgaria) (Kozlowsky, 1982), a raven bone with notches retrieved in the Micoquian layer of Zaskalnaya VI (Crimea) (Tsvelykh et al., 2014; Majkić et al., 2017) and a hyena femur with a set of incisions and a cave bear cervical vertebra showing subparellel marks respectively recovered in the Mousterian sites of Les Pradelles (France) (d'Errico et al., 2018) and Pešturina Cave (Serbia) (Majkić et al., 2018b). Also the flute of Divje Babe (Slovenia) has been included within the Neandertal artistic evidence (Turk, 1997; Chase and Nowell, 1998; Tuniz et al., 2012; Turk and Kosir, 2017).

During the Middle to Upper Palaeolithic transition complex thinking is well documented in Europe. The Chatelperronian and the Uluzzian are the techno-complexes to have yielded the most numerous findings, mainly ornaments. Various kinds of adornment objects have been retrieved in the Chatelperronian sites of Grotte du Renne (perforated bones and pierced teeth – Caron et al., 2011), Quincay (pierced teeth –Granger and Lévêque, 1997), Caune de Belvis (shells –Taborin, 1993) and Saint-Césaire (shells –Lévêque and Vandermeersch, 1980). Colouring substances were frequently used in order to obtain pigment powder at several Chatelperronian sites (Dayet et al., 2014). Further evidence of possible symbolic behavior are the engraved motifs adorning some bone implements from Grotte du Renne (d'Errico et al., 1998).
In Central Europe ornaments are curiously scanty as they are limited to a bone pendant and two pierced teeth retrieved at Bacho Kiro (Bulgaria) (Kozlowski, 1982), a perforated fossil gastropod from Willendorf II (Austria) (Felgenhauer,1956 and1959; Hahn,1993), and an ivory disc with a central hole, maybe a pendant, found at Ilsenhöhle (Germany) (Hulle, 1977).

The Uluzzian personal ornaments, mainly consisting of tusk shells, are distributed from Northern to Southern Italy up to Greece, attesting to a close cultural affinity among groups even over long distances. This pattern probably accounts for a common origin or an intimate ideological interaction (or both) within Uluzzian people. In other words, early MHs who arrived in Italy and in the Balkans had common technological and cultural traditions (see also Marciani et al. in this special issue) and were able to widely share their knowledge in the frame of a possible ethnic identity. In this light tusk shells could play the role of a cultural and social marker, as the use of ornaments, like other kinds of body modifications (tattoos, scarifications, ear piercing, lip and neck stretching etc..) is considered to be directly tied to the mental model the individual and the group represent themselves *vs* other individuals and/or groups (Boyd and Richerson, 1987; Nettle and Dunbar, 1997; McElreath et al., 2003; Newell et al.,1990;Vanhaeren and d'Errico, 2006).
The uniqueness of tusk shell phenomenon is underlined by the absence of this kind of ornaments in the IUP assemblages in general (Stiner et al, 2002; Campbell, 2017). Among the transitional techno-complexes, the occurrence of tusk shells in the Chatelperronian of Saint Césaire (Granger and Lévêque, 1997), is, so far, a singular exception.
The presence of coulouring substances is documented in some Uluzzian sites, even with decorative purposes as shown by the ochered tusk beads from Riparo Broion (Peresani et al., 2019a). This latter site is also the one to have yielded artistic evidence, an engraved stone.

In Europe, several Protoaurignacian sites, like La Louza (Taborin, 1993), Isturitz (Normand and Turq, 2005), L'Arbreda (Maroto et al., 1996; Soler Sublis et al., 2008) and Rotschild (in this site also fossil shells were used) (Bon, 2002; Sacchi, 1996; Taborin, 1993) yielded shell ornaments. The preference for basket-shaped shells is considered a distinctive feature of both the Protoaurignacian and the Aurignacian techno-complexes. This characteristic is reminiscent of the shell ornaments of the Early Ahmarian (Otte et al., 2011). Also pierced teeth (frequently fox and red-deer) are recorded in the assemblages of Rotschild (Bon, 2002; Sacchi, 1996; Taborin, 1993) and Isturitz (Normand and Turq, 2005) in France and Cueva Morin (González Echegaray and Freeman, 1971) in Spain.
In the Italian Protoaurignacian there is a vast assortment of ornamental taxa (mostly small size gastropods) with an evident change in the composition of personal adornments with respect to the Uluzzian. A significant role is played by *Homalopoma sanguineum*, possibly for its typical red coloration. A choice towards shells with red or yellow colorations has been noted for the Upper Palaeolithic in general, maybe because the high visibility of these colours may have had a peculiar symbolic significance (Álvarez-Fernández, 2006) or, more simply, for the attraction exercised by the colour itself. Concomitantly, tusk shells notably decrease and only a few specimens are recorded in Italian and European sites (Fiocchi, 1998; Bertola et al., 2013; Zilhão, 2007, Peresani et al., 2019b). This diversity between the Uluzzian and the Protoaurignacian ornamental suite could be traced back to their distinct ethnic identities, confirming, therefore, the different origins of these two techno-complexes.

## 5 Conclusions

The framework emerging between 50 ka and 36 ka cal. BP in Italy is characterized by an evolving situation, in which biological, cultural and technological (sometimes revolutionary) innovations occur. When considering this scenario in the light of the topics discussed above, some interesting remarks are possible. Firstly, we can note that the Mousterian patchy evidence regarded as symptomatic of behavioural modernity seems to be mostly related to the "intellectual" (otherwise mental or spiritual) sphere rather than to the technological field. The Mousterian actually displays some activities that do not seem directly related to subsistence needs. However, these appear as flash experiences, not shared by the most Neandertal groups. The punctuated nature of these findings is also suggestive of a sort of "inability/lack of interest" (possibly due to differently structured societies, different psychic and material needs etc...) in transmitting innovative conceptual information over time and space, which prevented, in facts, the sharing of these acquisitions. In other words, the occurrence in the Mousterian of formal bone tools, ornamental items, colouring substances and other "non-utilitarian" elements appears to be occasional even if sometimes controversial. This background does not consent to hypothesize a systematic achievement by Neandertals of those behaviours which are commonly considered as a prerogative of Early Upper Palaeolithic MH societies.

The Uluzzian is characterized by shared systematic modern behaviours in all its aspects: from lithic technology (see Marciani et al. in this special issue) to bone technology and ornaments. Among adornment objects, exclusively made on shells, tusks appear to be a sort of hallmark of the Uluzzian groups because of their number and distribution. The same uniformity is evoked by bone tools.

The Protoaurignacian, which differs from the Uluzzian especially as far as the ornamental kit is concerned, displays, even internally, some geographical differentiation in the amount and assortment of both ornaments and bone tools. Going south, the depletion of some typical features usually connected to ethnic identity is anthropologically consistent with the notion of a spread from the north of this techno-complex.

## Acknowledgements

This project has received funding from the European Research Council (ERC) under the European Union’s Horizon 2020 research and innovation programme (grant agreement No 724046); http://www.erc-success.eu/.

We thank the Soprintendenza Archeologia, Belle Arti e Paesaggio per le Province di Brindisi, Lecce e Taranto, for kindly supporting our research and fieldwork in Apulia over the years.

Special thanks are due to Professors Arturo Palma di Cesnola and Paolo Gambassini for giving us permission of studying materials from their excavations. We are grateful to Stefano Ricci for his help in editing figure 4.

The authors also thank the Soprintendenza Archeologia, Belle Arti e Paesaggio per la città metropolitana di Genova e le province di Imperia, Savona e La Spezia and the Polo Museale della Liguria for facilitating and supporting fieldwork in Liguria. Recent fieldwork at Bombrini was funded also by the Fonds Québécois pour la Recherche – Société et Culture (grant 2016-NP-193048) to J. Riel-Salvatore.
Research at Fumane is coordinated by the Ferrara University (M.P.) in the framework of a project supported by the Ministry of Culture - Veneto Archaeological Superintendency, public institutions (Lessinia Mountain Community - Regional Natural Park, Fumane Municipality, Veneto Region - Department for Cultural Heritage), and private associations and companies. Research at Riparo del Broion and Grotta di San Bernardino is designed by Ferrara University (M.P.) and was supported by MIBAC, the Province of Vicenza, the Veneto Region – Department for Cultural Heritage, and the Italian Ministry of Research and Education.

**Author contributions**

Conceptualization: Simona Arrighi.
Original draft: Simona Arrighi, Adriana Moroni.
Review & editing: Simona Arrighi, Adriana Moroni, Laura Tassoni, Francesco Boschin, Federica Badino, Eugenio Bortolini, Paolo Boscato, Jacopo Crezzini, Carla Figus, Manuela Forte, Federico Lugli, Giulia Marciani, Gregorio Oxilia, Fabio Negrino, Julien Riel-Salvatore, Matteo Romandini, Enza Elena Spinapolice, Marco Peresani, Annamaria Ronchitelli, Stefano Benazzi.

**References**

Alhaique, F., Cristiani, E., Molari, C., 2006. Un osso “lavorato” nel Musteriano di Grotta Breuil (Monte Circeo). Analisi delle modificazioni, sperimentazione e confronti. Atti della XXXIX Riunione Scientifica dell’Istituto Italiano di Preistoria e Protostoria, Istituto Italiano di Preistoria e Protostoria, Firenze, pp. 803-814.

Álvarez Fernández, E., 2006. Los objetos de adorno-colgantes del Paleolítico superior y del Mesolítico en la cornisa cantábrica y en el valle del Ebro: una visión europea. Ph. D. Thesis, University of Salamanca, Spain.

Armand, D., Delagnes, A., 1998. Les retouchoirs en os d’Artenac (couche 6c): perspectives archéozoologiques, taphonomiques et expérimentales. In: Brugal, J.P., Meignen, L., Patou-Mathis, M. (Eds.), Economie préhistorique: les comporte-ments de subsistance au Paléolithique, XVIIIe rencontres internationals d’Archéologie et d’Histoire d’Antibes. APDCA, Sophia Antipolis, pp. 193-203.

Bachechi, L., 2001. Un osso inciso da Grotta Costantini (Balzi Rossi, Italia). Archivio per l'Antropologia e l'Etnologia CXXXI, 65-79.

Badino, F., Pini R., Ravazzi, C., Margaritora, D., Arrighi, S., Bortolini, E., Figus, C., Giaccio, B., Lugli, F., Marciani, G., Monegato, G., Moroni, A., Negrino, F., Oxilia, G., Peresani, M., Romandini, M., Ronchitelli, A., Spinapolice, E. E., Zerboni, A., Benazzi, S.. An overview of Alpine and Mediterranean palaeogeography, terrestrial ecosystems and climate history during MIS 3 with focus on the Middle to Upper Palaeolithic transition. Submitted to Quaternary International, this special issue.

Bahn, P., 1996. New developments in Pleistocene art. Evolutionary Anthropology 4, 204-215.

Barge H., 1983. Essai sur les parures du Palèolithique supérieur dans le Sud de la France. La faune malacologique aurignacienne de l'Abri Rothschild (Cabrières, Hérault). Bullettin du Musée d'Anthropologie Préhistorique de Monaco 27, 69-83.

Bartolomei, G., Broglio, A., Cattani, L., Cremaschi, M., Guerreschi, A., Mantovani, E., Peretto, C., Sala, B., 1982. I depositi würmiani del Riparo Tagliente. Annali dell'Università di Ferrara 3, n.s., sez.15, 61-105.

Bar-Yosef, O., 2003 Reflections on the selected issues of the Upper Palaeolithic. In: Goring-Morris, A. N., Belfer-Cohen, A. (Eds.), More than Meets the Eye: Studies on Upper Palaeolithic. Diversity in the Near East, Oxbow Books, Oxford, pp. 265-273.

Bar-Yosef, O., Bordes, J.G., 2010. Who were the makers of the Châtelperronian culture? Journal of Human Evolution 59, 586-593. https://doi.org/10.1016/j.jhevol.2010.06.009.

Bar-Yosef Mayer, D., Vandermeersch B., Bar-Yosef, O., 2009. Shells and ochre in Middle Paleolithic Qafzeh Cave, Israel: indications for modern behavior. Journal of Human Evolution 56, 307-314. https://doi.org/10.1016/j.jhevol.2008.10.005.

Bednarik, R. G., 1995. Concept-mediated marking in the Lower Palaeolithic, Current Anthropology 36, 605-16. https://doi.org/10.1086/204406.

Bednarik, R. G., 2006. The Middle Paleolithic engravings from Oldisleben, Germany. L'Anthropologie XLIV (2), 113-121.

Benazzi, S., Douka, K., Fornai, C., Bauer, C.C., Kullmer, O., Svoboda, J., Pap, I., Mallegni, F., Bayle, P., Coquerelle, M., Condemi, S., Ronchitelli, A., Harvati, K., Weber, K. W., 2011. Early dispersal of modern humans in Europe and implications for Neanderthal behaviour. Nature 479, 525-528. https://doi.org/10.1038/nature10617.

Benjamin, J., Rovere, A., Fontana, A., Furlani, S., Vacchi, M., Inglis, R.H.H., Galili, E., Antonioli, F., Sivan, D., Miko, S., Mourtzas, N., Felja, I., Meredith-Williams, M., Goodman-Tchernov, B., Kolaiti, E., Anzidei, M., Gehrels, R., 2017. Late Quaternary sea-level changes and early human societies in the central and eastern Mediterranean Basin: An interdisciplinary review. Quaternary International 449, 29–57. doi: 10.1016/j.quaint.2017.06.025.

Benini, A., Boscato, P., Gambassini, P., 1997. Grotta della Cala (Salerno): industrie litiche e faune uluzziane ed aurignaziane. Rivista di Scienze Preistoriche XLVIII, 37-95.

Bertola, S., Peresani, M., Peretto, C., Thun-Hohenstein, U., 1999. Le site paléolithique moyen de la Grotta della Ghiacciaia (Préalpes de Vénétie, Italie du Nord). L'Anthropologie 103 (3), 377-390.

Bertola, S., Broglio, A., Cristiani, E., De Stefani, M., Gurioli, F., Negrino, F., Romandini, M., Vanhaeren, M., 2013. La diffusione del primo Aurignaziano a sud dell'arco alpino. Preistoria Alpina 47, 123-152.

Biddittu, I., Cassoli, P., Malpieri, L., 1967. Stazione musteriana in Valle Radice nel comune di Sora. Quaternaria XI, 321-348.

Bietti, A., Negrino, F., 2008. L'Aurignacien et le Gravettien du Riparo Mochi, l'Aurignacien du Riparo Bombrini: comparaisons et nouvelles perspectives. Archives de l'Institut de Paléontologie Humaine 39, 133-140.

Blasco, R., Rosell, J., Cuartero, F., Fernández Peris, J., Gopher, A., Barkai, R., 2013. Using bones to shape stones: MIS 9 bone retouchers at both edges of the Mediterranean Sea. PlosOne 8 (10), e76780. https://doi.org/10.1371/journal.pone.0076780.

Bon, F., 2002. L'Aurignacien entre mer et océan. Réflexion sur l'unité des phases anciennes de l'Aurignacien dans le sud de la France. Société Préhistorique Française, Paris.

Bordes, F., 1969. Os percé moustérien et os gravé acheuléen du Pech de l'Azé II. Quaternaria 11, 1-6.

Borgia, V., Boschin, F., Ronchitelli, A., 2016. Bone and antler working at Grotta Paglicci (Rignano Garganico, Foggia, southern Italy). Quaternary International 403, 23-39. https://doi.org/10.1016/j.quaint.2015.11.116.

Borgia, V., Crezzini, J., Ronchitelli, A., 2017. Grotta del Cavallo (LE): strumento in osso inedito dell'Uluzziano antico. In: Radina, F. (Ed.), Preistoria e Protostoria della Puglia, Studi di Preistoria e Protostoria 4, Istituto Italiano di Preistoria e Protostoria, Firenze, pp. 663-667.

Borzatti von Löwenstern, E., Magaldi, D., 1967. Ultime ricerche nella Grotta dell'Alto (S. Caterina di Nardò). Rivista di Scienze Preistoriche XXII (II), 205-250.

Boyd, R., Richerson, P. J., 1987. The evolution of ethnic markers. Cultural Anthropology 2, 65-79. https://doi.org/10.1525/can.1987.2.1.02a00070.

Broglio, A., Angelucci, D. E., Peresani, M., Lemorini, C., Rossetti, P., 1996. L'industrie Protoaurignacienne de la Grotta di Fumane: données préliminaires. Actes du XIIIe Congrès de l'UISPP. Forlì, Italy, pp. 495-509.

Broglio, A., De Stefani, M., Gurioli, F., Peresani, M., 2006. Les peintures aurignaciennes de la Grotte de Fumane (Monts Lessini, Préalpes de la Vénétie). INORA 44, 1-8.

Bronk Ramsey, C., 2017. Methods for Summarizing Radiocarbon Datasets. Radiocarbon 59, 1809–1833. https://doi.org/10.1017/RDC.2017.108.

Brooks, A.S., Helgren, D.M., Cramer, J.S., Franklin, A., Hornyak, W., Keating, J.M., Klein, R.G., Rink, W.J., Schwarcz, H., Leith Smith, J.N., Stewart, K., Todd, N.E., Verniers, J., Yellen, J.E., 1995. Dating and context of three Middle Stone Age sites with bone points in the Upper Semliki Valley, Zaire. Science 268(5210), 548-553.

Brooks, A.S., Yellen, J. E., Potts, R., Behrensmeyer, A. K., Deino, A. L., Leslie, D. E., Ambrose, S. H., Ferguson, J. R., d'Errico, F., Zipkin, A. M., Whittaker, S., Post, J., Veatch, E. G., Foecke, K., Clark, J.B., 2018. Long-distance stone transport and pigment use in the earliest Middle Stone Age. Science 360 (6384), 90-94. https://doi.org/10.1126/science.aao2646.

Campbell, G., 2017. The Reproduction of Small Prehistoric Tusk Shell Beads. In: Bar-Yosef Mayer, D., Bonsall, C., Choyke, A.M. (Eds.), Not Just for Show: The Archaeology of Beads, Beadwork and Personal Ornaments. Oxbow Books, Oxford and Philadelphia, pp.168-224.

Caron, F., d'Errico, F., Del Moral, P., Santos, F., Zilhão, J., 2011. The reality of Neandertal symbolic behavior at the grotte du Renne, Arcy-sur-Cure, France. PlosOne 6 (6), e21545. http://dx.doi.org/10.1371/journal.pone.0021545.

Cavallo, G., Fontana, F., Gonzato, F., Peresani, M., Riccardi, M.P., Zorzin, R., 2017. Textural, microstructural and compositional characteristics of Fe-based geomaterials and Upper Palaeolithic ocher in the Lessini Mountains, Northeast Italy: implications for provenance studies. Geoarchaeology 32 (4), 437-455. https://doi.org/10.1002/gea.21617.

Cavallo, G., Fontana, F., Gialanella, S., Gonzato, F., Riccardi, M.P., Zorzin, R., Peresani, M., 2018. Heat treatment of mineral pigment during the Upper Palaeolithic in North-East Italy. Archaeometry 60 (5), 1045–1061. https://doi.org/10.1111/arcm.12360.

Chase, P. G., Nowell, A., 1998. Taphonomy of a suggested Middle Paleolithic bone flute from Slovenia. Current Anthropology 39, 549–553.

Chella, P., 2002. I manufatti in steatite. In: Campana, N., Maggi, R. (Eds.), Archeologia in Valle Lagorara. Diecimila anni di storia intorno a una cava di diaspro. Origines. Istituto Italiano di Preistoria e Protostoria, Firenze, pp. 265-280.

Churchill, S.H., Smith, F.H., 2000. Makers of the Early Aurignacian. American Journal of Physical Anthropology 31, 61-115. https://doi.org/10.1002/1096-8644(2000)43:31

Crémades, M., 1996. L'expression graphique au Paléolithique inférieur et moyen: l'example de l'Abri Suard (La-Chaise-de-Vouthon, Charente). Bullettin de la Socété Préhistorique Française 93 (4), 494-501.

Daujeard, C., Moncel, M-H., Fiore, I., Tagliacozzo, A., Bindon, P., Raynal, J-P., 2014. Middle Paleolithic bone retouchers in Southeastern France: Variability and functionality. Quaternary International 326-327, 492-518. https://doi.org/10.1016/j.quaint.2013.12.022.

Dayet, L, d’Errico, F., Garcia-Moreno, R., 2014. Searching for consistencies in Châtelperronian pigment use. Journal of Archaeological Science 44, 180-193. https://doi.org/10.1016/j.jas.2014.01.032.

Dayet, L., Le Bourdonnec, F-X., Daniel, F., Porraz, G., Texier, P-J., 2015. Ochre Provenance and Procurement Strategies During the Middle Stone Age at Diepkloof Rock Shelter, South Africa. Archaeometry 58 (5), 807-829. https://doi.org/10.1111/arcm.12202.

d'Errico, F., Villa P., 1997. Holes and grooves: the contribution of microscopy and taphonomy to the problem of art origins. Journal of Human Evolution 33, 1-31. https://doi.org/10.1006/jhev.1997.

d’Errico, F., Henshilwood, C.S., 2007. Additional evidence for bone technology in the southern African Middle Stone Age. Journal of Human Evolution 52 (2), 142-163. https://doi.org/10.1016/j.jhevol.2006.08.003.

d'Errico, F., Zilhao, J., Baffier, D., Julien, M., Pelegrin, J., 1998. Neandertal acculturation in Western Europe? A critical review of the evidence and its interpretation. Current Anthropology 39, 1-44. https://doi.org/10.1086/204689.

d’Errico, F., Vanhaeren, M., Wadley, L., 2008. Possible shell beads from the Middle Stone Age layers of Sibudu Cave, South Africa. Journal of Archaeological Science 35 (10), 2675-2685. https://doi.org/10.1016/j.jas.2008.04.023.

d’Errico, F., Vanhaeren, M., Barton, N., Bouzouggar, A., Mienis, H., Ritcher, D., Hublin, J-J., McPherron, S.P., Lozouet, P., 2009. Additional evidence on the use of personal ornaments in the Middle Palaeolithic of North Africa. Proceedings of the National Academy of Science of the United States of America 106 (38), 16501-16506. https://doi.org/10.1073/pnas.0903532106.

d'Errico, F., Backwell, L. R.,Wadley, L., 2012a. Identifying regional variability in Middle Stone Age bone technology: The case of Sibudu Cave. Journal of Archaeological Science 39 (7), 2479-2495. https://doi.org/10.1016/j.jas.2012.01.040.

d'Errico, F., Borgia, V., Ronchitelli, A., 2012b. Uluzzian bone technology and its implications for the origin of behavioural modernity. Quaternary International 259 (9), 59-71. https://doi.org/10.1016/j.quaint.2011.03.039

d'Errico, F., Doyon, L., Colagé, I., Queffelec, A., Le Vraux, E., Giacobini, G., Vandermeersch, B., Maureille, B., 2018. From number sense to number symbols. An archaeological perspective. Philosophical Transactions of the Royal Society B: Biological Sciences 373. https://doi.org/10.1098/rstb.2016.0518

Eren, M.I., Diez-Martin, F., Dominguez-Rodrigo, M., 2013. An empirical test of the relative frequency of bipolar reduction in Beds VI, V, and III at Mumba Rockshelter, Tanzania: implications for the East African Middle to Late Stone Age transition. Journal of Archaeological Science 40 (81), 248-256. https://doi.org/10.1016/j.jas.2012.08.012.

Felgenhauer, F., 1956-1959. Willendorf in der Wachau. Monographie der Paläolith-Fundstellen I-VII. Mitteilungen der Prähistorischen Kommission der Österreichischen, Akademie der Wissenschaften VIII–IX, Rohrer, Wien.

Fiocchi, C., 1998. Contributo alla conoscenza del comportamento simbolico di *Homo sapiens sapiens*. Le conchiglie marine nei siti del Paleolitico superiore europeo: strategie di approvvigionamento, reti di scambio, utilizzo. Ph.D. Thesis, Universities of Bologna, Ferrara and Parma, Italy.

Freund, G., 1987. Das Paläolithikum der Oberneder-Höhle (Landkreis Kelheim/Donau). Quartär Bibliothek 5 (1), 210-221.

Gambassini, P., 1997. Il Paleolitico di Castelcivita. Culture e ambiente. Electa, Napoli.

Gernone, G., Maggi, R., 1998. Lavorazione della steatite alla Pianaccia di Suvero (Alta Val di Vara, La Spezia). In: Nicolis, F., Mottes, E. (Eds.), Simbolo ed enigma. Il bicchiere campaniforme e l'Italia nella preistoria europea del III millennio a.C. Provincia Autonoma di Trento, Servizio Beni Culturali, Ufficio Beni Archeologici, Trento, pp. 95-97.

González Echegaray, J., Freeman, L. G., 1971. Cueva Morin: Excavaciones 1966–68. Patronato de las Cuevas Prehistoricas de la Provincia de Santander, Santander.

Goren-Inbar, N., 1986. A Figurine from the Acheulian Site of Berekhat Ram. Mitekufat Haeven: Journal of the Israel Prehistoric Society, 7-12.

Glen, E., Kaczanowski, K., 1982. Human remains. In: Kozłowski, J. K. (Ed.), Excavation in the Temnata Cave, Final Report, PWN, Warszawa, pp. 75-80.

Granger, J.M., Lévêque, F., 1997. Parure castelperronienne et aurignacienne: étude inédites de dents percées et comparison. Comptes Rendus de l'Académie des Sciences - Series IIA - Earth and Planetary Science 325 (7), 537-543. https://doi.org/10.1016/S1251-8050(97)89874-6

Gravina, B., Bachellerie, F., Caux, S., Discamps, E., Faivre, J-P., Galland, A., Michel, A., Teyssandier, N., Bordes, J-G., 2018. No Reliable Evidence for a Neanderthal-Châtelperronian Association at La Roche-à-Pierrot, Saint-Césaire. Scientific Reports 8,15134. https://doi.org/10.1038/s41598-018-33084-9.

Hahn, J., 1976. Der Sirgenstein, eine urgeschichtliche Höhlenstation im Achtal. Kulturdenkmale in Baden-Württemberg. Kleine Führer, 24, Stuttgart.

Hahn, J., 1993. L'origine du Paléolithique supérieur en Europe centrale: les datations C14. In: Cabrera, V. (Ed.), El origen del hombre moderno en el suroeste de Europa, Universidad Nacional de Educación a Distancia, Madrid, pp. 61-80.

Henshilwood, C.S., 2012. Late Pleistocene Techno-traditions in Southern Africa: A Review of the Still Bay and Howiesons Poort, c. 75–59 ka. Journal of World Prehistory 25, 205-237. https://doi.org/10.1007/s10963-012-9060-3.

Henshilwood, C.S., d'Errico, F., Marean, C. W, Milo, G., Yates, R., 2001. An early bone tool industry from the Middle Stone Age at Blombos Cave, South Africa: implications for the origins of modern human behaviour, symbolism and language. Journal of Human Evolution 41 (6), 631-678. https://doi.org/10.1006/jhev.2001.0515.

Higham, T., Jacobi, R., Julien, M., David, F., Basell, L., Wood, R., Davies, W., Bronk Ramsey, C., 2010. Chronology of the Grotte du Renne (France) and implications for the context of ornaments and human remains within the Châtelperronian. Proceedings of the National Academy of Science of the United States of America 107 (47), 20234-20239. https://doi.org/10.1073/pnas.1007963107.

Higham, T., Brock, F., Bronk Ramsey, C., Davies, W., Wood, R., Basell, L., 2011. Chronology of the site of Grotte du Renne, Arcy-sur-Cure, France: implications for Neanderthal symbolic behaviour. Before Farming 2011 (2), 1-9.https://doi.org/10.3828/bfarm.2011.2.1

Hoffecker, J. F., 2009. The spread of modern humans in Europe. Proceedings of the National Academy of Science of the United States of America 106 (38), 16040–16045. https://doi.org/10.1073/pnas.0903446106.

Hoffmann, D. L., Standish, C. D., Pike, A. W. G, García-Diez, M., Pettitt, P. B., Angelucci, D. E., Villaverde, V., Zapata, J., Milton, J. A., Alcolea-González, J., Cantalejo-Duarte, P., Collado, H., de Balbín, R., Lorblanchet, M., Ramos-Muñoz, J., Weniger, G-C., Zilhão, J., 2018a. Dates for Neanderthal art and symbolic behaviour are reliable. Nature Ecology & Evolution 2, 1044-1045. https://doi.org/10.1038/s41559-018-0598-z.

Hoffmann, D. L., Standish, C. D., García-Diez, M., Pettitt, P.B., Milton, J. A., Zilhão, J., Alcolea-González, J. J., Cantalejo-Duarte, P., Collado, H., de Balbín, R., Lorblanchet, M., Ramos-Muñoz, J., Weniger, G.-Ch., Pike, A. W. G., 2018b. U-Th dating of carbonate crusts reveals Neandertal origin of Iberian cave art. Science 359, 912-915. https://doi.org/10.1126/science.aap7778.

Holt, B., Negrino, F., Riel-Salvatore, J., Formicola, V., Arellano, A., Arobba, D., Boschian, G., Churchill, S.E., Cristiani, E., Di Canzio, E., Vicino, G., 2018. The Middle-Upper Paleolithic transition in Northwest Italy: new evidence from Riparo Bombrini (Balzi Rossi, Liguria, Italy). Quaternary International. https://doi.org/10.1016/J.QUAINT.2018.11.032.

Hublin, J-J., 2013. The Makers of the Early Upper Paleolithic in Western Eurasia. In: Smith, F.H., Ahern J.C.M. (Eds), The Origins of Modern Humans: Biology Reconsidered, John Wiley & Sons, Hoboken, NJ, pp. 223-252.

Hublin, J.-J., Talamo, S., Julien, M., David, F., Connet, N., Bodu, P., Vandermeersch, B., Richards, M.P., 2012. Radiocarbon dates from the Grotte du Renne and Saint Césaire support a Neandertal origin for the Châtelperronian. Proceedings of the National Academy of Science of the United States of America 109 (46), 18743-18748. https://doi.org/10.1073/pnas.1212924109.

Hülle, W. M., 1977. Die Ilsenhöhle unter Burg Ranis/Thüringen. Eine paläolitische Jägerstation, Gustav Fischer, Stuttgart.

Jacobs, Z., Duller, G. A. T., Wintle, A.G., Henshilwood, C.S., 2006. Extending the chronology of deposits at Blombos Cave, South Africa, back to 140 ka using optical dating of single and multiple grains of quartz. Journal of Human Evolution 51 (3), 255-273. https://doi.org/10.1016/j.jhevol.2006.03.007.

Jacobs, Z., Roberts, R.G., Galbraith, R.F., Deacon, H.J., Grün, R., Mackay, A., Mitchell, P., Vogelsang, R., Wadle, L., 2008. Ages for the middle stone age of Southern Africa: Implications for human behavior and dispersal. Science 322 (5902), 733–735. https://doi.org/10.1126/science.1162219.

Jéquier, C. A., 2014. Techno-economy of hard osseous materials between Middle and Upper Palaeolithic in Norhern Italy and Slovenia. Ph.D. Thesis, University of Ferrara, Italy.

Jéquier, C. A., Romandini, M., Peresani, M., 2012. Les retouchoirs en matières dures animales : une comparaison entre Moustérien final et Uluzzien. Comptes Rendus Palevol 11, 283-292. https://doi.org/10.1016/j.crpv.2011.12.001.

Jéquier, C., Peresani, M., Romandini, M., Delpiano, D., Joannes-Boyau, R., Lembo, G., Livraghi, A., López-García, J.M., Obradović, M., Nicosia, C., 2015. The De Nadale Cave, a single layered Quina Mousterian site in the North of Italy. Quartär 62, 7-21. https://doi.org/10.7485/QU62_1.

Jéquier, C., Peresani, M., Livraghi, A., Romandini, M., 2018. Same but different: 20,000 years of bone retouchers from Northern Italy. A multi-cultural approach from Neanderthals to Anatomically Modern Humans. In: Hutson, J.M., García-Moreno, A., Noack, E.S., Turner, E., Villaluenga, A., Gaudzinski-Windheuser, S. (Eds.), The origins of bone tool technologies. Neuwied, Monrepos, pp. 269-286.

Joordens, J. C. A., d'Errico, F., Wesselingh, F. P., Munro, S., de Vos, J., Wallinga, J., Ankjærgaard, C., Reimann, T., Wijbrans, J. R., Kuiper, K. F., Mücher, H. J., Coqueugniot, H., Prié, V., Joosten, I., van Os, B., Schulp, A. S., Panuel, M., van der Haas, V., Lustenhouwer, W., Reijmer, J. J. G., Roebroeks, W., 2014. Homo erectus at Trinil on Java used shells for tool production and engraving. Nature 518, 228-231. https://doi.org/10.1038/nature13962.

Julien, M., Baffier, D., Liolios, D., 2002. L'outillage en matières dures animals. In: Schmider, B. (Ed.), L'Aurignacien de la grotte du Renne. Les fouilles d'André Leroi-Gourhan à Arcy-sur-Cure (Yonne). CNRS Éditions, Paris, pp. 217-250.

Julien, M-A., Hardy, B., Stahlschmidt, M. C., Urban, B., Serangeli, J., Conard, N. J., 2015. Characterizing the Lower Paleolithic bone industry from Schöningen 12 II: A multi-proxy study. Journal of Human Evolution 69, 264-286. http://dx.doi.org/10.1016/j.jhevol.2015.10.006.

Klein, R.G., 1999. The Human Career: Human Biological and Cultural Origins.The University of Chicago Press, Chicago.

Kozłowski, J., 1982. Excavation in the Bacho Kiro Cave (Bulgaria): Final Report. Polish Scientific Publishers, Warsaw.

Kuckenburg, M., 2001. Als der Mensch zum Schpfer wurde: An den Wurzeln der Kultur. Klett Cotta, Stuttgart.

Kuhn, S.L., Stiner, M.C., 1992. New research at Riparo Mochi, Balzi Rossi (Liguria): preliminary results. Quaternaria Nova II, 77-90.

Kuhn, S.L., Stiner, M.C., 1998. The earliest aurignacian of Riparo Mochi (Liguria, Italy). Current Anthropology 39 (3), 175-189.

Leonardi, P., 1975. Incisioni pre-leptolitiche europee. Annali dell'Università di Ferrara, n.s. XV (II), 283-321.

Leonardi, P., 1979. Una serie di ritoccatoi prevalentemente musteriani del RiparoTagliente in Valpantena presso Verona. Preistoria Alpina 15, 7-15.

Leonardi, P., 1980. Incisioni e ossa lavorate musteriane del Riparo Tagliente a Stallavena nei Monti Lessini presso Verona. Bullettino di Paletnologia Italiana, n.s. XXIV, 7-18.

Leonardi, P., 1981. Raschiatoio musteriano del Riparo Solinas di Fumane (Verona) con incisioni sul cortice. Atti dell'Accademia Roveretana degli Agiati di Rovereto VI (XX), 87-89.

Leonardi, P., 1983. Incisioni musteriane del Riparo Tagliente in Valpantena nei Monti Lessini presso Verona (Italia). Homenaje al prof. Al magro Basch. Ministerio de Cultura, Madrid.

Leonardi, P., 1988. Art palèolithique et mobilier en Italie. L'Anthropologie 92, 139-201.

Leroi-Gourhan, A., 1961. Les fouilles d'Arcy-sur-Cure (Yonne). Gallia Préhistoire IV (2), 3-16.

Leroi-Gourhan, A., 1964. Les religions de la Préhistoire (Paléolithique). Presses Universitaires de France. Paris.

Lévêque, F., Vandermeersch, B., 1980. Découverte de restes humains dans un niveau castelperronien à Saint-Césare (Charente-Maritime). Comptes Rendus de l'Académie des Sciences Paris 291 (2), 187-189.

McBrearty, S., Brooks, A.S., 2000. The revolution that wasn't: a new interpretation of the origin of modern human behavior. Journal of Human Evolution 59 (5), 453-563. https://doi.org/10.1006/jhev.2000.0435.

McElreath, R., Boyd, R., Richerson, P. J., 2003. Shared norms and the evolution of ethnic markers. Current Anthropology 44 (1), 122-129. https://doi.org/10.1086/345689.

Majkić, A., Evans, S., Stepanchuk, V., Tsvelykh, A., d'Errico, F., 2017. A decorated raven bone from the Zaskalnaya VI (Kolosovskaya) Neanderthal site, Crimea. PlosOne 12 (3), e0173435. https://doi.org/10.1371/journal.pone.0173435

Majkić, A., d'Errico, F., Stepanchuk, V., 2018a. Assessing the significance of Palaeolithic engraved cortexes. A case study from the Mousterian site of Kiik-Koba, Crimea. PlosOne 13 (5), e0195049. https://doi.org/10.1371/journal.pone.0195049.

Majkić, A., d'Errico, F., Milošević, S., Mihailović, D., Dimitrijević, V., 2018b. Sequential Incisions on a Cave Bear Bone from the Middle Paleolithic of Pešturina Cave, Serbia. Journal of Archaeological Method and Theory 25, 69–116. https://doi.org/10.1007/s10816-017-9331-5.

Malerba, G., Giacobini, G., 1996. Les retouchoirs sur éclats diaphysaires du Paléolithique moyen et supérieur de trois sites de l'Italie nord orientale (Grotte de San Bernardino, Abri de Fumane et Abri Tagliente). In: Patou-Mathis, M. (Ed.), Actes du XIIIème Congrès UISPP, Volume 6/1, Workshop 4: L'industrie sur os du Paléolithiqu einférieur et moyen: nouvelles méthodes d'analyse, Forlì, Italy, pp. 167-171.

Mania, D., Mania, U., 1988. Deliberate engravings on bone artefacts of Homo Erectus. Rock Art Research 5, 91-97.

Marciani, G., Ronchitelli, A., Arrighi, S., Badino, F., Bortolini, E., Boscato, P., Boschin, F., Crezzini, J., Delpiano, D., Falcucci, A., Figus, C., Lugli, F., Negrino, F., Oxilia, G., Romandini, M., Riel-Salvatore, J., Spinapolice, E. E., Peresani, M., Moroni, A., Benazzi, S. Lithic techno-complexes in Italy from 50 to 39 thousand years BP: an overview of cultural and technological changes across the Middle-Upper Palaeolithic boundary. Submitted to Quaternary International, this special issue.

Maroto, J., Soler, N., Fullola, J. M., 1996. Cultural change between Middle and Upper Paleolithic in Catalonia. In: Carbonell, E., Vaquero, M. (Eds.), The Last Neandertals, the First Anatomicallv Modern Hurnans: a Tale about the Human Diversity. Cultural Change and Human Revolution at 40 ka BP. Universitat Rovira i Virgili, Tarragona, pp. 210-250.

Marshack, A., 1976. Some implications of the Paleolithic symbolic evidence for the origin of language. Current Anthropology 17 (2), 274-282.

Martini, F., Sarti, L., Buggiani, S., 2004. Incisioni su pietra da Grotta del Cavallo (Lecce): contributo al dibattito sulle esperienze grafiche neandertaliane. Rivista di Scienze Preistoriche LIV, 271-289.

Mellars, P.A., 1973. The character of the Middle-Upper Paleolithic transition in south-west France. In: Renfrew, C. (Ed.), The Explanation of Culture Change: Models in Prehistory, Duckworth, London, pp. 255-276.

Mellars, P., 2005. The impossible coincidence. A single-species model for the origins of modern human behavior in Europe. Evolutionary Anthropology 14 (1), 12-27. https://doi.org/10.1002/evan.20037.

Mellars, P., 2006. Archaeology and the dispersal of modern humans in Europe: Deconstructing the “Aurignacian.” Evolutionary Anthropology 15, 167–182. https://doi.org/10.1002/evan.20103.

Mellars, P., 2011. The earliest modern humans in Europe. Nature 479 (7374), 483-485. https://doi.org/10.1038/479483a.

Moigne, A.-M., Valensi, P., Auguste, P., Garcia-Solano, J., Tuffreau, A., Lamotte, A., Barroso, C., Moncel, M.-H., 2016. Bone retouchers from Lower Palaeolithic sites: Terra Amata, Orgnac 3, Cagny l'Epinette and Cueva del Angel. Quaternary International 409, 195-212. https://doi.org/10.1016/j.quaint.2015.06.059.

Moncel, M-H., Chiotti, L., Gaillard, C., Onoratini, G., Pleurdeau, D., 2012. Non-utilitarian lithic objects from the European Paleolithic. Archaeology Ethnology & Anthropology of Eurasia 40 (1), 24-40. https://doi.org/10.1016/j.aeae.2012.05.004.

Moog, F., 1939. Paläolithische Freilandstation im Älteren Löß von Wyhlen (Amt Lörrach). Badische Fundberichte 15, 36-52.

Moroni, A., Boscato, P., Ronchitelli, A., 2013. What roots for the Uluzzian? Modern behaviour in Central-Southern Italy and hypotheses on AMH dispersal routes. Quaternary International 316, 27-44. https://doi.org/10.1016/j.quaint.2012.10.051.

Moroni, A., Ronchitelli, A., Arrighi, S., Aureli, D., Bailey, S. E., Boscato, P., Boschin, F., Capecchi, G., Crezzini, J., Douka, K., Marciani, G., Panetta, D., Ranaldo, F., Ricci, S., Scaramucci, S., Spagnolo, V., Benazzi, S., Gambassini, P., 2018. Grotta del Cavallo (Apulia – Southern Italy). The Uluzzian in the mirror. Journal of Anthropological Sciences 96, 1-36. https://doi.org/10.4436/jass96004.

Morin, E., Laroulandie, V., 2012. Presumed symbolic use by diurnal raptors by Neanderthals. PlosOne 7 (3), e32856. https://doi.org/10.1371/journal.pone.0032856.

Mussi, M., 1990. Le peuplement de l’Italie à la fin du Paléolithique moyen et au début du Paléolithique supérieur. In: Fairzy, C. (Ed.), Paléolithique Moyen Récent et Paléolithique Supérieur

Ancien en Europe. Ruptures et Traditions: Examen Critique des Documents Archéologiques. Memoires du Musée de Préhistoir d'Ile-de France, vol. 3, Centre National de Documentation Pédagogique, Paris, pp. 251-262.

Normand, C., Turq, A., 2005. L'Aurignacien de la grotte d'Isturitz (France): la production lamellaire dans la séquence de la salle Saint-Martin. In: Le Brun-Ricalens, F., Bordes, J.-G., Bon, F. (Eds.), Productions lamellaires attribuées à l'Aurignacien: châınes opératoires et perspectives technoculturelles. ArchéoLogiques 1, Musée National d'Histoire et d'Art, Luxembourg, pp. 375-394.

Negrino, F., Colombo, M., Cremaschi, M., Serradimigni, M., Tozzi, C., Ghiretti, A. 2017. Estese officine litiche del Paleolitico medio-superiore sui rilievi appenninici di Monte Lama-Castellaccio-Pràrbera (Bardi, Parma). Studi di Preistoria e Protostoria 3. Istituto Italiano di Preistoria e Protostoria, Firenze, pp. 59-68.

Nettle, D., Dunbar, R. I. M., 1997. Social markers and the evolution of reciprocal exchange. Current Anthropology 38 (1), 93-99. https://doi.org/10.1086/204588.

Newell, R. R., Kielman, D., Constandse-Westermann, T. S., Van der Sanden, W. A., Van Gijn, A., 1990. An inquiry into the ethnic resolution of Mesolithic regional groups: The study of their decorative ornaments in time and space. Brill, Leiden.

Ortega Cobos, D., Soler, N., Maroto, J., 2005. La production de lamelles pendant l'Aurignacien archaïque dans la grotte de l'Arbreda (Espagne): organisation de la production, variabilité des méthodes et objectifs. In: Le Brun-Ricalens, F., Bordes, J.-G., Bon, F. (Eds.), Productions lamellaires attribuées à l'Aurignacien : chaînes opératoires et perspectives technoculturelles, Actes du XIVeCongrès de l'UISPP, Liège, Belgium, pp. 359-374.

Otte, M., Shidrang, S.,  Zwyns, N., Flas, D., 2011. New radiocarbon dates for the Zagros Aurignacian from Yafteh cave, Iran. Journal of Human Evolution 61 (3), 340-346. https://doi.org/10.1016/j.jhevol.2011.05.011.

Palma di Cesnola, A., 1966. Il Paleolitico superiore arcaico (facies uluzziana) della Grotta del Cavallo, Lecce (continuazione). Rivista di Scienze Preistoriche 21, 3-59.

Palma di Cesnola, A., 1989. L'Uluzzien: faciès italien du Leptolithique archaïque. L'Anthropologie 93, 783-812.

Palma di Cesnola, A., 1993. Il Paleolitico superiore in Italia. Garlatti e Razzai, Firenze.

Patou-Mathis, M., 2002. Retouchoirs, compresseurs, percuteurs . . .Os à impressions et éraillures. Société Préhistorique Française, Paris.

Peresani, M., Fiore, I., Gala, M., Romandini, M., Tagliacozzo, A., 2011. Late Neandertals and the intentional removal of feathers as evidenced from bird bone taphonomy at Fumane Cave 44 ky B.P.,

Italy. Proceedings of the National Academy of Science of the United States of America,108 (10) 3888-3893; https://doi.org/10.1073/pnas.1016212108.

Peresani, M., Vanhaeren, M., Quaggiotto, E., Queffelec, A., d'Errico, F. 2013. An Ochered Fossil Marine Shell From the Mousterian of Fumane Cave, Italy. PlosOne 8 (7), e68572. https://doi.org/10.1371/journal.pone.0068572.

Peresani, M., Romandini, M., Duches, R., Jéquier, C., Nannini, N., Pastoors, A., Picin, A., Schmidt, I., Vaquero, M., Weniger, G-C., 2014a. New evidence for the Mousterian and Gravettian at Rio Secco Cave, Italy. Journal of Field Archaeology 39 (4), 401-416. https://doi.org/10.1179/0093469014Z.00000000098.

Peresani, M., Dallatorre, S., Astuti, P., Dal Colle, M., Ziggiotti, S., Peretto, C., 2014b. Symbolic or utilitarian? Juggling interpretations of Neanderthal behavior: new inferences from the study of engraved stone surfaces. Journal of Anthropological Sciences 92, 233-255. https://doi.org/10.4436/JASS.92007.

Peresani, M., Cristiani, E., Romandini, M., 2016. The Uluzzian technology of Grotta di Fumane and its implication for reconstructing cultural dynamics in the Middle-Upper Palaeolithic transition of Western Eurasia. Journal of Human Evolution 91, 36-56. https://doi.org/10.1016/j.jhevol.2015.10.012.

Peresani, M., Bertola, S., Delpiano, D., Benazzi, S., Romandini, M., 2019a. The Uluzzian in north Italy: insights around the new evidence at Riparo del Broion. Archaeological and Anthropological Sciences. https://doi.org/10.1007/s12520-018-0770-z.

Peresani, M., Forte, M., Quaggiotto, E., Colonese, A. C., Romandini, M., Cilli, C., Giacobini G., 2019b. Marine shell exploitation in the Early Upper Palaeolithic. Re-examination of the shell assemblages from Fumane Cave (NE Italy). PaleoAnthropology, 64-81. https://doi.org/10.4207/PA.2019.ART124

Pitti, C., Sorrentino, C., Tozzi, C., 1976. L'industria di tipo Paleolitico superiore arcaico della Grotta La Fabbrica (Grosseto). Nota preliminare. Atti della Società Toscana di Scienze Naturali, Memorie, Serie A 83, 174-201.

Radovčić, D., Sršen, A. O., Radovčić, J., Frayer, D.W., 2015. Evidence for Neandertal Jewelry: Modified White-Tailed Eagle Claws at Krapina. PlosOne 10 (3), e0119802. https://doi.org/10.1371/journal.pone.0119802.

Raynal, J. P., Séguy R., 1986. Os incisé aucheléen de Saint Anne I (Polignac-Haute Loire). Revue Archéologique du Centre de la France 25, 79-80.

Reimer, P. J., Bard, E., Bayliss, A., Beck, J. W., Blackwell, P. G., Ramsey, C. B., Buck, C. E., Cheng, H., Edwards, R. L., Friedrich, M., Grootes, P. M., Guilderson, T. P., Haflidason, H., Hajdas, I., Hatté, C., Heaton, T. J., Hoffmann, D. L., Hogg, A. G., Hughen, K. A., Kaiser, K. F., Kromer,

B., Manning, S. W., Niu, M., Reimer, R. W., Richards, D. A., Scott, E. M., Southon, J. R., Staff, R. A., Turney, C. S. M., van der Plicht, J., 2013. IntCal13 and Marine13 Radiocarbon Age Calibration Curves 0–50,000 Years cal BP. Radiocarbon 55, 1869-1887. https://doi.org/10.2458/azu_js_rc.55.16947.

Riel Salvatore, J., Negrino, F., 2018. Human adaptations to drastic climatic change in Liguria across the Middle-Upper Paleolithic transition. Journal of Quaternary Science 33 (3), 313-322. https://doi.org/10.1002/jqs.3005.

Rigaud, A., 2007. Retouchoirs sur éclats diaphysaires ou “affutoirs” de Labastide (Hautes-Pyrénées). Archéologie des Pyrénées occidentales et des Landes 26,193-200.

Rodríguez-Vidal, J., d’Errico, F., Giles Pacheco, F., Blasco, R., Rosell, J., Jennings, R. P., Queffelec, A., Finlayson, G., Fa, D. A., Gutiérrez López, J. M., Carrión, J. S., Negro, J. J., Finlayson, S., Cáceres, L. M., Bernal, M. A., Fernández Jiménez, S., Finlayson, C., 2014. A rock engraving made by Neanderthals in Gibraltar. Proceedings of the National Academy of Science of the United States of America 111 (37) 13301-13306. https://doi.org/10.1073/pnas.1411529111.

Ronchitelli, A., Boscato, P., Gambassini, P., 2009. Gli ultimi Neandertaliani in Italia: aspetti culturali. In: Facchini, F., Belcastro, G. (Eds.), La lunga storia di Neandertal. Biologia e comportamento, Jaka Book, Bologna, pp. 257-288.

Romandini, M., Cristiani, E., Peresani, E., 2014a. A retouched bone shaft from the Late Mousterian at Fumane cave (Italy). Technological, experimental and micro-wear analysis. Comptes Rendus Palevol 14 (1), 63-72. https://doi.org/10.1016/j.crpv.2014.08.001.

Romandini, M., Peresani, M., Laroulandie, V., Metz, L., Pastoors, A., Vaquero, M., Slimak, L., 2014b. Convergent Evidence of Eagle Talons Used by Late Neanderthals in Europe: A Further Assessment on Symbolism. PlosOne 9 (7), e101278. https://doi.org/10.1371/journal.pone.0101278.

Romandini, M., Terlato, G., Nannini, N., Tagliacozzo, A., Benazzi, S., Peresani, M., 2018. Bears and humans, a Neanderthal tale. Reconstructing uncommon behaviors from zooarchaeological evidence in southern Europe. Journal of Archaeological Science 90, 71-91. https://doi.org/10.1016/j.jas.2017.12.004.

Romandini, M., Crezzini, J., Bortolini, E., Boscato, P., Boschin F., Carrera, L., Tagliacozzo A., Terlato, G., Arrighi, S., Badino, F., Figus C., Lugli, F., Marciani, G., Oxilia, G., Moroni, A., Negrino, F., Peresani, M., Riel-Salvatore J., Ronchitelli A., Spinapolice, E. E., Benazzi, S. Macromammal and bird assemblages of Late Middle to Upper Palaeolithic transition in Italy.Submitted to Quaternary International, this special issue.

Rougier, H., Crevecoeur, I., Beauval, C., Posth, C., Flas, D., Wißing, C., Furtwängler, A., Germonpré, M., Gómez-Olivencia, A., Semal, P., van der Plicht, J., Bocherens, H., Kraus, J., 2016. Neandertal cannibalism and Neandertal bones used as tools in Northern Europe**.** Scientific Reports, 6, 29005. https://doi.org/10.1038/srep29005.

Sacchi, D., 1996. Le Paléolithique supérieur en Pyrénées et en Languedoc méditerranéen (1991–1996). In: Otte, M. (Ed.), Le Paléolithique Supérieur Européen: Bilan Quinquennal 1991–1996. Études et Recherches Archéologiques de l'Université de Liége 76, Liège, pp. 269-283.

Serangeli, J., van Kolfschoten, T., Starkovich, B.M., Verheijen, I., 2015. The European saber-toothed cat (Homotheriumlatidens) found in the "Spear Horizon" at Schöningen (Germany). Journal of Human Evolution 89, 172-180. https://doi.org/10.1016/j.jhevol.2015.08.005.

Siret, M. L., 1925. Emploi de l'os dans la retouche des silex moustériens. Bulletin de la Société Préhistorique Française 22 (5), 208-210.

Slimak, L., Fietzke, J., Geneste, J-M., Ontañón, R., 2018. Comment on "U-Th dating of carbonate crusts reveals Neandertal origin of Iberian cave art". Science 361 (6408), eaau1371. https://doi.org/10.1126/science.aau1371.

Soler Subils, J., Soler Masferrer, N., Maroto, J., 2008. L'Arbreda's archaic Aurignacian dates clarified. Eurasian Prehistory 5 (2), 45- 55.

Soressi, M., McPherron, S. P., Lenoir, M., Dogandžić, T., Goldberg, P., Jacobs, Z., Maigrot, Y., Martisius, N. L., Miller, C. E., Rendu, W., Richards, M., Skinner, M. M., Steele, T. E., Talamo, S., Texier, J-P., Neandertals made the first specialized bone tools in Europe. Proceedings of the National Academy of Science of the United States of America 110 (35), 14186-14190. https://doi.org/10.1073/pnas.1302730110.

Soulier, M.-S., Goutas, N., Normand, C., Legrand, A., White, R., 2014. Regards croisés de l'archéozoologue et du technologue sur l'exploitation des ressources animales à l'Aurignacien archaïque: l'exemple d'Isturitz (Pyrénées-Atlantiques, France). In: Thiébault, C., Claud, É., Costamagno, S. (Eds.), Exploitation des ressources organiques à la fin du Paléolithique moyen et au début du Paléolithique supérieur: interactions entre environnement et comportements techniques, Actes de la session E du XXVIIe Congrès Préhistorique de France, Société Préhistorique Française, Paris, pp. 315-332.

Spennato, A.G., 1981. I livelli protoaurignaziani della Grotta di Serra Cicora (Nardò-Lecce). Studi per l'Ecologia del Quaternario 3, 61-76.

Stiner, M., 1999. Trends in Paleolithic mollusc exploitation at Riparo Mochi (Balzi Rossi, Italy): Food and ornaments from the Aurignacian through Epigravettian. Antiquity 73, 735-754.

Stiner, M. C., 2010. Shell ornaments from the Upper Paleolithic through Mesolithic layers of Klissoura Cave 1 by Prosymna (Peloponnese, Greece). Eurasian Prehistory 7 (2), 287-308.

Stiner, M. C., Pehlevan, C., Sağir, M., Ŏzer, I., 2002. Zooarchaeological studies at Üçağızlı Cave: Preliminary results on Paleolithic subsistence and shell ornaments. Araşturma Sonuçları Toplantısı 17, 29-36.

Taborin, Y., 1993. La parure en coquillage au Paléolithique. XXIX Suppléments à Gallia Préhistoire), CNRS éd, Paris.

Tartar É., 2015 - Origine et développement de la technologie osseuses aurignacienne en Europe occidentale: bilan des connaissances actuelles. In: White, R., Bourrillon, R., Bon, F. (Eds.), Aurignacian Genius: art, technologie et société des premiers hommes modernesen Europe. Actes du symposium international, New York University, P@lethnologie 7, pp. 34-56.

Tejero, J-M., 2014. Towards complexity in osseous raw material exploitation by the first anatomically modern humans in Europe: Aurignacian antler working. Journal of Anthropological Archaeology 36, 72-92. http://dx.doi.org/10.1016/j.jaa.2014.08.004.

Tejero, J-M., Grimaldi, S., 2015. Assessing bone and antler exploitation at Riparo Mochi (Balzi Rossi, Italy): implications for the characterization of the Aurignacian in South-western Europe. Journal of Archaeological Science 61, 59-77. http://dx.doi.org/10.1016/j.jas.2015.05.003.

Teyssandier, N., Liolios, D., 2008. Le concept d'Aurignacien : entre rupture préhistorique et obstacle épistémologique. Bulletin de la Société Préhistorique Français 105, 737-747. https://doi.org/10.3406/bspf.2008.13782.

Thun Hoehnstein, U., Bertolini, M., 2012. Strategie di sussistenza e sfruttamento delle risorse animali. In: Peretto, C. (Ed.). L'insediamento Musteriano di Grotta Reali, Rocchetta a Volturno, Molise, Italia. Annali dell'Università di Ferrara, Sez. Museologia Scientifica e Naturalistica, pp.27-34.

Thun Hohenstein, U., Bertolini, M., Channarayapatna, S., Modolo, M., Peretto, C., 2018. Bone retouchers from two north italian Middle Palaeolithic sites: Riparo Tagliente and Grotta della Ghiacciaia, Verona, In: Hutson, J.M., García-Moreno, A., Noack, E.S., Turner, E., Villaluenga, A., Gaudzinski-Windheuser, S. (Eds.), The origins of bone tool technologies. Neuwied, Monrepos, pp. 235-250.

Tozzi, C., 1974. L'industria musteriana della Grotta di Gosto sulla montagna di Cetona (Siena). Rivista di Scienze Preistoriche XXIX (2), 271-304.

Tuniz, C., Bernardini, F., Turk, I., Dimkaroski, L., Mancini, L., Dreossi, D., 2012. Did Neanderthals play music? X-ray computed micro-tomography of the Divje babe 'flute'. Archaeometry 54, 581-590. https://doi.org/10.1111/j.1475-4754.2011.00630.x.

Turk, I., 1997. Mousterian 'bone flute' and other finds from Divje babe I cave site in Slovenia. Založba ZRC, Ljubljana.

Turk, M., Kosir, A., 2017. Mousterian osseous artefacts? The case of Divje babe I, Slovenia. Quaternary International 450, 103-115. https://doi.org/10.1016/j.quaint.2016.12.012.

Tromnau, G., 1983. Ein Mammutknochen-Faustkeil aus Rhede, Kreis Borken (Westfalen). Archäologisches Korrespondenzblatt 13 (3), 287-289.

Tsvelykh, A. N., Stepanchuk, V.N., 2014. An artefact made of a bird bone from Zaskalnaya VI (Kolosovskaya) Mousterian site in Crimea. Zamyatninskiysbornik 3,124–127, (In Russian).

Vandevelde, S., Brochier, J. E., Petit, C., Slimak, L., 2017. Establishment of occupation chronicles in Grotte Mandrin using sooted concretions: Rethinking the Middle to Upper Paleolithic transition. Journal of Human Evolution 112, 70-78. https://doi.org/10.1016/j.jhevol.2017.07.016.

Vanhaeren, M., d’Errico, F., 2006. Aurignacian ethno-linguistic geography of Europe revealed by personal ornaments. Journal of Archaeological Science 33, 1105-1128. https://doi.org/10.1016/j.jas.2005.11.017.

Vanhaeren, M, d’Errico, F., Stringer, C., James, S. L., Todd, J. A., Mienis, H. K., 2006. Middle Palaeolithic shell beads in Israel and Algeria. Science 312, 1785-1787. https://doi.org/10.1126/science.1128139.

Vanhaeren, M., d’Errico, F., Van Niekerk, K. L., Henshilwood, C. S., Erasmus, R. M., 2013. Thinking strings: additional evidence for personal ornament use in the Middle Stone Age at Blombos Cave, South Africa. Journal of Human Evolution 64 (6), 500-517. https://doi.org/10.1016/j.jhevol.2013.02.001.

van Kolfschoten, T., Parfitt, S.A., Serangeli, J., Bello, S.M., 2015. Lower Paleolithic bone tools from the ‘Spear Horizon’ at Schöningen (Germany). Journal of Human Evolution 89, 226-263. https://doi.org/10.1016/j.jhevol.2015.09.012.

Villa, P., d'Errico, F., 2001. Bone and ivory points in the Lower and Middle Paleolithic of Europe. Journal of Human Evolution 41 (2), 69-112. https://doi.org/10.1006/jhev.2001.0479.

Villa, P., Pollarolo, L., Conforti, J., Marra, F., Biagioni, C., Degano, I., Lucejko, J. J., Tozzi, C., Pennacchioni, M., Zanchetta, G., Nicosia, C., Martini, M., Sibilia, E., Panzeri, L., 2018. From Neandertals to modern humans: new data on the Uluzzian. PlosOne, 13 (5), 0196786-0196842. https://doi.org/10.1371/journal.pone.0196786.

Vincent, A., 1993. L’outillage osseux au Paleolithique moyen: une nouvelle approche. Ph. D. Thesis, University of Paris X-Nanterre, France.

Wadley, L., 2003. How some archaeologists recognize culturally modern behaviour. South African Journal of Science 99 (5–6), 247-250.

Wadley, L., 2005. Putting ochre to the test: replication studies of adhesives that may have been used for hafting tools in the Middle Stone Age. Journal of Human Evolution 49 (5), 587-601. https://doi.org/10.1016/j.jhevol.2005.06.007.

White, R., 2001. Personal Ornaments from the Grotte du Renne at Arcy-sur-Cure. Athena Review 2, 41-46.

Wolpoff, M. L., 1996. Neandertals of the Upper Paleolithic. In: Carbonell, E., Vaquero, M. (Eds.), The Last Neandertals, the First Anatomically Modern Humans. Fundacio Catalana per la Recerca, Tarragona, pp. 51–76.

Yellen, J.E., 1998. Barbed bone points: tradition and continuity in Saharan and sub-Saharan Africa. African Archaeological Review15.173-198. https://doi.org/10.1023/A:1021659928822.

Zilhão, J., 2007. The Emergence of Ornaments and Art: An Archaeological Perspective on the Origins of "Behavioral Modernity". Journal of Archaeological Research 15,1-54. http://dx.doi.org/10.1007/s10814-006-9008-1.

Zilhão, J., Angelucci, D. E., Badal-García, E., d'Errico, F., Daniel, F., Dayet, L., Douka, K., Higham, T. H. G., Martínez-Sánchez, M. J., Montes-Bernárdez, R., Murcia-Mascarós, S., Pérez-Sirvent, C., Roldán-García, C., Vanhaeren, M., Villaverde, V., Wood, R., Zapata, J., 2010. Symbolic use of marine shells and mineral pigments by Iberian Neandertals. . Proceedings of the National Academy of Science of the United States of America 107 (3), 1023-1028. https://doi.org /doi/10.1073/pnas.0914088107.

Zilhão, J., Bank, W. E., d'Errico, F., Gioia, P., 2015. Analysis of Site Formation and Assemblage Integrity Does Not Support Attribution of the Uluzzian to Modern Humans at Grotta del Cavallo. PlosOne 10 (7), e0131181. https://doi.org /10.1371/journal.pone.0131181.

**Figure captions**

Fig.1 Localization of the MIS3 Italian sites yielding bone tools and/or ornaments and other non utilitarian items. The Italian Peninsula shows a sea level of 70 m below the present-day coastline, based on the global sea-level curve (Benjamin et al., 2017) but lacking the estimation of post-MIS3 sedimentary thickness and eustatic magnitude (sketch map courtesy by S. Ricci, University of Siena).

Fig. 2 Mousterian bone retouchers from Grotta di Fumane showing percussion traces: *Cervus elaphus* metacarpal and close-up of the percussion traces (layer A6) (1). Double retoucher made from *Alces* or *Megalocers* tibia and close-up of the percussion traces (layers A5+A6) (2).

Fig. 3 Uluzzian bone tools. Grotta del Cavallo (layers EIII, EII-I, D) (1-- 8. Grotta della Cala (layer D14) (9). Grotta di Castelcivita (layers "rsi", "rpi", "rsa") (10-14). Grotta La Fabbrica (layer 2) (15). Riparo Broion (layer 1g) (16-19). Grotta di Fumane (layer A3) (20-21) (Modified after d'Errico et al., 2012b; Peresani et al., 2016, Villa et al., 2018; Peresani et al., 2019a).

Fig. 4 Protoaurignacian bone tools. Riparo Bombrini (layers A3-A1): pointed tool (1), fragmentary tool (2) needle (4) and awls fragments (3, 5-7); Grotta di Fumane (layers A2-A1): split based point

recovered at the top of layer A1(8), awls (9-10) and distal portion of a needle or a awl; Grotta Paglicci (layer 24): awl (12); Grotta della Cala (layers 12-13) fragmentary bone points (11-14).

Fig.5 Eagle claws with cut marks from Riosecco and Fumane and close-up of the anthropic signs indicating their intentional removal (1). The *Aspa marginata* recovered at Grotta Fumane and zoom on the striations on the inner lip. The striations are consistent with the presence of a thread, attesting the use of the shell as a pendant (2) (Modified after Peresani et al., 2013).

Fig. 6 Uluzzian ornamental assemblages. Riparo Broion: *Antalis vulgaris* (1-2 and 4-5), *Antalis inaequicostata* (3), *Theodoxus danubialis* (6). Grotta della Cala: *Antalis vulgaris* (7-8), *Glycimeris nummaria* (15-16), *Homalopoma sanguineum* (9), *Clanculus corallinus* (10). Grotta del Cavallo: *Tritia neritea* (17), *Columbella rustica* (18), *Antalis* sp. (19-30).

Fig.7 Ornaments and artistic items from Italian Portoaurignacian sites. Grotta di Fumane: sample of the ornamental shells, *Tritia mutabils* (1), *Homalopoma sanguineum* (2), *Tritia pellucida* (3), *Glyicimeris nummaria* (4), teeth pendants (5-6), engraved rib from a medium-sized ungulate (7). Riparo Bombrini: sample of the ornamental shells (8-10), worked steatite fragments (11-15), fragmentary steatite pendant (16), bird bone with notches and incisions (17). Grotta della Cala: sample of ornamental shells, *Homalopoma sanguineum* (18-23). Grotta di Castelcivita: sample of ornamental shells, *Pecten jacobeus* (24) *Homalopoma sanguineum* (25)